\documentclass[times]{elsarticle}

\usepackage{moreverb}

\usepackage[colorlinks,bookmarksopen,bookmarksnumbered,citecolor=red,urlcolor=red]{hyperref}

\newcommand\BibTeX{{\rmfamily B\kern-.05em \textsc{i\kern-.025em b}\kern-.08em
T\kern-.1667em\lower.7ex\hbox{E}\kern-.125emX}}

\usepackage{todonotes}
\usepackage{listings}
\usepackage{amsmath}
\usepackage{amssymb}
\usepackage{pbox}
\usepackage{xspace}
\usepackage{paralist}
\usepackage{xcolor}
\definecolor{light-gray}{gray}{0.9}
\definecolor{dark-gray}{gray}{0.6}

\usepackage{enumitem}

\usepackage{algpseudocode}
\usepackage{algorithm}
\usepackage{eqparbox}

\newcommand{\tool}{$ABT$\xspace}
\newcommand{\newtool}{$ABT_{2.0}$\xspace}
\newcommand{\app}{$ERP$\xspace}

\newenvironment{change}{}{}

\newenvironment{change2}{}{}

\usepackage{changepage}
\usepackage{tcolorbox}

\begin{document}

\title{On Introducing Automatic Test Case Generation in Practice: A Success Story and Lessons Learned}

\author{Matteo Brunetto$^*$, Giovanni Denaro$^*$, Leonardo Mariani$^*$, Mauro Pezz\'e$^{+,\#}$}

\address{$^*$Universit\`a di Milano-Bicocca, Milano, Italy --- $^+$USI Universit\`a della Svizzera Italiana, Lugano, Switzerland --- $^{\#}$SIT Schaffhausen Institute of Technology, Schaffhausen, Switzerland}

\address{[matteo.brunetto $|$ giovanni.denaro $|$ leonardo.mariani]@unimib.it ---  mauro.pezze@usi.ch}

\begin{abstract}
The level 
\begin{change2}
and quality
\end{change2}of automation dramatically affects software testing activities, determines costs and effectiveness of the testing process, and largely impacts on the quality of the final product.  
While costs and benefits of automating many testing activities in industrial practice (including managing the quality process, executing large test suites, and managing regression test suites) are well understood and documented, the benefits and obstacles of automatically generating system test suites in industrial practice are not well reported yet, despite the recent progresses of automated test case generation tools.
Proprietary tools for automatically generating test cases are becoming common practice in large software organisations, and commercial tools are becoming available for some application domains and testing levels.
However, generating system test cases in small and medium-size software companies is still largely a manual, inefficient and ad-hoc activity.

This paper reports our experience in introducing techniques for automatically generating system test suites in a medium-size company. We describe the technical and organisational obstacles that we faced when introducing automatic test case generation in the development process of the company, and present the solutions that we successfully experienced in that context.
In particular, the paper discusses the problems of automating the generation of test cases by referring to a customised ERP application that the medium-size company developed for a third party multinational company, and presents \newtool, the test case generator that we developed by tailoring \tool, a research state-of-the-art GUI test generator, to their industrial environment. 
This paper presents the new features of \newtool, and discusses how these new features address the issues that we faced. 

\end{abstract}

\maketitle

\section{Introduction}

Designing effective test suites requires a significant effort that dramatically affects both testing and  development costs, with a relevant impact on the software development schedule, the project deadlines, and ultimately the quality of the final product.
\begin{change}The recent advances in automated test case generation and the new tools for automatically generating test suites can reduce the effort required to generate effective test suites, and thus improve the cost-effectiveness of test case generation activities. \end{change}

Several popular  research prototypes 
(for instance, Randoop~\cite{Pacheco:Randoop:ICSE:2007}, JBSE~\cite{Braione:enhancing:esecfse:2013}, Evosuite~\cite{Fraser:whole:2013} and SUSHI~\cite{,Braione:SUSHI:ISSTA:2017})
and commercial products 
(for instance, PEX~\cite{Tillmann:pex:TAP:2008} and Parasoft Testing C/C++ Test~\cite{Parasoft:C-C++Test:Web})
provide good support for automatically generating \emph{unit} test suites.  
The increasing success of test generation tools for unit testing already produced studies about the issues associated with their adoption in industry~\cite{Fraser:UnitTestsEmpirical:TOSEM:2015}.  

\begin{change2}
Tools for generating \emph{system} test cases can address applications in both critical~\cite{wang:automated:iciest:2018}
and non-critical domains~\cite{Almasi:industria:icse-seip2017,Arcuri:Evomaster:tosem:2019}.
The industrial practice for automatically generating system test cases for GUI applications offers tools that capture, generalize, and execute behaviors observed when users interact with the system~\cite{tosca}.    
\end{change2}
Studies about the effectiveness of research prototype tools for generating system test cases provide evidence of the effectiveness of interesting approaches on medium-scale and small-scale open source software systems,
notably the work on ABT~\cite{Mariani:GUI:STVR:2014}, Exsyst~\cite{Gross:Exsyst:ISSTA:2012}, WATEG~\cite{Thummalapenta:GuidedTestWeb:ICSE:2013}, GUITAR~\cite{Yuan:StateFeedback:TSE:2010}, and Gazoo~\cite{arlt:GAZOO:ISSRE:2012}.

\begin{change2}
However, introducing automated system testing solutions available in the form of research prototypes in industrial software development processes\end{change2} challenges automated test case generators with problems that can be hardly experienced with medium-scale and small-scale open source software~\cite{Braione:TestGenerator:SwQuality:2014}. 
\begin{change2}Some experiences reported so far refer to the successful introduction of automatic test case generation tools for unit testing~\cite{Almasi:industria:icse-seip2017} and REST API testing in industrial settings~\cite{Arcuri:Evomaster:tosem:2019}, but there is little attention to the specific issues that must be addressed to effectively introduce automatic GUI test case generators within commercial organizations. 
\end{change2}

\begin{change}
This paper reports an exploratory study that investigates the issues that arise when introducing a leading-edge-research GUI test generator, \tool~\cite{Mariani:Autoblacktest:ICST:2012}, in a commercial organisation. 
\tool generates system test cases by sampling possible GUI interaction sequences. \tool relies on Q-Learning, a machine learning algorithm, to steer the test generation process towards relevant functionalities of the target application~\cite{Sutton:IRL:1998}. 
\end{change}
In this paper we
\begin{inparaenum} [(i)]
\item introduce the relevant issues that led a software house to consider an automated approach for generating system test suites, 
\item present the sometimes unexpected technical and 
organisational obstacles that prevented the straightforward adoption of 
\begin{change}
\tool
\end{change}
for automatically generating system test suites, 
\item discuss the technical improvements that we made to overcome those obstacles, and that led to \newtool, a system test case generator \begin{change} that extends \tool to address\end{change} 
the specific industrial process, 
and
\item illustrate the lessons that we learned with our experience and that may generalise to other commercial applications that share the characteristics of our study.
\end{inparaenum}

The \begin{change} study\end{change} reported in this paper is the result of a one-year \begin{change}pilot\end{change} technology-transfer project.
\begin{change} 
It follows an exploratory design 
because of the initial lack of  clear propositions and hypotheses on the potential constraints that could limit the effectiveness of \tool in the considered industrial setting. 
\end{change}
During the project, we faced several challenges and learned several lessons 
about \emph{scalability}, \emph{test reporting} and \emph{oracles} that are specific to the problem of automatically generating test cases, and that represent major barriers to the adoption of automated system test case generation tools. 
Interestingly, we also found that we could largely or totally address the most relevant issues with domain-tailored solutions that adapt well to the specific business oriented applications considered in our project, thus reducing the need for difficult-to-find general solutions that would have significantly delayed the adoption of the test generator. 

In the light of the issues that we observed during the project, we extended the tool \tool with new solutions that exploit the domain specific characteristics of the applications under test, to produce an effective test generation tool, \newtool, tailored for testing business oriented applications.
Based on the results of our experiments with an ERP application considered in our project,  \begin{change}the study reported in\end{change} this paper provides \begin{change}initial\end{change} empirical evidence that our extensions are the key ingredients to turn \begin{change}our\end{change} academic test case generation tool \begin{change}\tool\end{change} from an ineffective to an effective solution for testing industrial business oriented applications. 
Our experience refers to a specific system testing technology and a specific product under test, \begin{change} and cannot be directly generalised. However,\end{change}
the challenges that we faced and the solutions that we designed can be generally relevant for both researchers and practitioners, who might think to apply similar strategies in different contexts to make automatic system testing more effective.

In a nutshell, this paper contributes to the state of the art by: 
 \begin{itemize}
 \item identifying some important challenges to automatically generating system test suites in the context of industrial software systems;
 \item proposing original extensions of the test generator \tool that improve the effectiveness of the approach for automatic system testing of industrial business oriented applications. \begin{change}While the core algorithm of \tool refers to our previous work~\cite{Mariani:Autoblacktest:ICST:2012}, the extensions reported in this paper are entirely novel contributions that we present for first time (Sections~\ref{sec:strategy} and~\ref{sec:oracles}). \end{change}
The proposed extensions 
  improve the scalability of the GUI exploration strategy of \tool,  produce effective test reports during the test generation process, and support the definition of domain specific test oracles;
 \item presenting the experience of exploiting the proposed extensions in the context of a case study concerned with generating tests for a business oriented application developed by our industrial partner.
 \end{itemize}

This paper is organized as follows. \begin{change}Section~\ref{sec:abt} overviews \tool, the technology that we identified as a viable test case generation tool for the target industrial case.\end{change}
Section~\ref{sec:challenges} discusses the main challenges in introducing automatic test case generation in an industrial development process by referring to our experience with a medium-size company that produces software solutions on demand. The section describes the industrial context of the technology transfer project, the challenges that we faced, and the domain-specific opportunities offered by the application under test.
Sections~\ref{sec:strategy} and~\ref{sec:oracles} discuss	the solutions to the technical and organisational challenges, respectively. These solutions led to \newtool, and to the extension of \tool to business oriented applications that meet the industrial requirements of our project.  The sections present the new features, and discuss experimental evidence of their effectiveness.
 Section~\ref{sec:results} distills some key lessons learned that can be useful to researchers and practitioners working on automatic system testing. \begin{change2}Section~\ref{sec:threats} discusses the main threats to validity.\end{change2} 
 Section~\ref{sec:related} surveys the related research efforts, and Section~\ref{sec:conclusions} summarizes the contributions of the paper and outlines our current research plans.  \section{AutoBlackTest}
\label{sec:abt}

AutoBlackTest (\tool) is the test generation technology that we referred to in our project.  In the first part of \begin{change}our exploratory case study\end{change}, we used \tool to gain sample data on the extent of the applicability and the limitations of a representative state-of-the-art test generator, when challenged with the industrial applications considered in the \begin{change}pilot\end{change} project. \begin{change}Later in the project, we used \tool\end{change} as a  platform to concretely develop and experience with new test generation strategies that we designed to exploit the opportunities identified in the project.
Our initial experience with introducing \tool in the industrial context was the driver that enlightens both the challenges and the opportunities that we discuss in  Sections~\ref{subsec:challenges}, which led to \newtool that enriches \tool with new functionalities to meet both technical and organisational industrial requirements. 
In this section we introduce \tool to make this paper self-contained. The interested readers can refer to~\cite{Mariani:Autoblacktest:ICSE:2011,Mariani:Autoblacktest:ICST:2012,Mariani:GUI:STVR:2014} for a comprehensive discussion of \tool.

\tool generates system test cases for applications that rely on GUI driven interactions based on Q-Learning~\cite{Sutton:IRL:1998}, a learning algorithm that \tool uses to steer the testing activity towards the most relevant functionalities of an application. 
In this paper, we refer to the \tool Q-Learning exploration strategy as \emph{Reinforcement Learning based Strategy (RLS)}.
Q-Learning is extensively used to address the problem of agents that learn how to interact with unknown environments. 
Q-learning agents learn how to interact with the environment by iteratively selecting and executing  actions, and updating a model of the system that estimates the utility of the actions according to the reactions of the environment.
Q-learning agents 
interleave random and model-based selection of actions, to take advantage of the incrementally build models, and increasingly explore relevant portions of the environments.   

The \tool Q-learning agent computes the utility of actions as the impact of the actions on the GUI, based on the intuition that the impact of an action on the state of the application should produce observable effects. 
Actions that produce relevant unseen transitions in the GUI, for instance actions that lead to successfully submitting a complete registration form, 
likely correspond to important interactions, and are thus given high utility values. While actions that produce negligible or already seen transitions, for instance actions that repetitively lead to incomplete forms or to error windows that bring the application back to its initial state, likely correspond to scarcely relevant interactions, and are thus given low utility values. 
By rewarding actions based on their impact on the GUI, the \tool Q-learning agent steers the testing activity towards combinations of actions that correspond to important interactions, and reduces the amount of repetitive actions with respect to selecting actions with a purely random process.

The \tool testing process initializes the Q-learning model to the home page of the application, and builds interactions (i.e., test cases) as sequences of episodes, where the number of  sequences and the length of each sequence are parameters. 
Each episode starts from a random state of the current Q-learning model, and executes a fixed number of actions that are selected according to the $\epsilon$-greedy policy.\footnote{ Q-Learning supports several policies,  the $\epsilon$-greedy policy is demonstrated to be effective for test case generation~\cite{Mariani:GUI:STVR:2014}.} 
The $\epsilon$-greedy policy alternates exploration and exploitation. Exploration selects actions never executed before, exploitation executes the most useful action according to the knowledge accumulated so far by the agent. 
In particular, when \tool is in a state not in the Q-learning model yet, the policy selects a random action (exploration). 
When \tool is in a state already in the Q-learning model, the policy executes a random action with probability $\epsilon$ (exploration) and the action with the highest Q-value according to the model with probability $1-\epsilon$ (exploitation), where $\epsilon$ is a user-provided parameter. 

Executing actions may require some data, for instance, editing a textfield requires identifying the data to be entered in the field. 
\tool determines values by using a catalog of values that associates labels, such as \texttt{email} and \texttt{date}, to a set of values, such as \texttt{mariani@disco.unimib.it} and \texttt{20-04-2019}. When interacting with an input widget, \tool analyzes the GUI to determine the label of the GUI that better describes the values expected by the input widget~\cite{Becce:Widget:FASE:2012}. For instance, \tool can determine that an email address should be entered in an input field from the presence of the label \texttt{email} next to the input widget. When entering data, \tool randomly selects a value from the set of values associated with the label, \texttt{email} in the example. When the label is not present in the catalog, \tool selects a string from a default set of values. 

Software applications may include complex functionalities that are hard to successfully execute using random exploration only. For instance, an application may include a form that can be successfully submitted only by filling several fields. An execution that enters values in all the fields of the form and then submits the form has little probability to be produced randomly. To address these cases, \tool supports the execution of complex actions, which are short workflows that execute a sequence of actions according to a specific strategy. For instance, when an application displays a large form, a complex action may facilitate submitting the form by entering a value in each field and then clicking the submit button. 

\tool considers the execution of complex actions before applying the standard \mbox{$\epsilon$-greedy} policy. In particular, if the current state of the application allows to execute a complex action, \tool selects it with probability $\textit{pcomplex}$. If the complex action is not executed, \tool applies the regular \mbox{$\epsilon$-greedy} policy.

\tool tool is implemented in Java and integrates IBM Functional Tester~\cite{IBM:FuncTester:WEBSITE}, a commercial capture and replay tool, to interact with the GUI of an application. \tool can support the same GUI frameworks supported by IBM Functional Tester. The experience reported in this paper exploits the support for .NET applications. \tool also uses the free library TeachingBox~\cite{Ertel:TeachingBox:ICAR:2009} to handle the Q-learning model.

In our industrial experience, we executed \tool using the $\epsilon$-greedy policy with \mbox{$\epsilon=0.7$} and with \mbox{$\textit{pcomplex}=0.5$}. 
We initialized the catalog of input values with \mbox{35 entries} comprised of valid user names, emails and data item identifiers, for instance, the identifier of an invoice, to satisfy the input fields with domain specific constraints, and used random strings for the unconstrained input fields. 
The testing process consisted of \mbox{50 episodes}, each one composed of 30 actions. \begin{change2}We determined the values of the parameters based on both our experience with the tool~\cite{Mariani:GUI:STVR:2014} and a trial and error fine-tuning process executed on the target industrial application.\end{change2}

In the next sections, we discuss how we adapted 
\begin{change}\tool to face the challenges of business oriented applications, \end{change} 
ending up with extending \tool to \newtool  by redesigning the GUI exploration strategy and extending its test reporting capabilities. 
 \section{Testing Business Oriented Applications: Challenges and Strategies}
\label{sec:challenges}

In a recent joint project between the LTA research laboratory\footnote{LTA -- Laboratory for software Testing and Analysis, Universit\`a degli studi di Milano Bicocca} and an industrial partner\footnote{A medium company producing customized software solutions on demand operating in the area of North Italy, whose identity is not disclosed for non-disclosure agreement policies}, we evaluated the benefits of  automating the generation of system test suites for business applications that rely on GUI-driven interactions. 
In this section we report the technical obstacles that derive from the complexity of the target application, and the organizational issues due to the specificity of the software process of the industrial partner.
We discuss obstacles and opportunities offered by the distinctive nature of the business oriented application targeted in the project, by focusing on scenarios that are common to many diverse applications and domains, and that can be exploited beyond the scope of our project to effectively automate the generation of system test cases.  
We introduce the business oriented application that we targeted in the project, by presenting an essential overview and discussing challenges and opportunities for automatically generating system test suites, effective in the business context.

\subsection{Business Oriented Applications: A Sample}
\label{sec:challenges:subject}

We conducted our \begin{change}exploratory \end{change} study on an application that our industrial partner selected as representative of typical customized software applications developed on demand for third parties. 
In this section, we outline the distinctive characteristics of the selected application,  which we refer to as \app here on. In a nutshell, \app  is an application that handles the commercial process of a company by managing data entities like orders, invoices, shipping activities, and maintenance requests. 
We illustrate our industrial experience using a simplified GUI with the same characteristics of the GUI of the original application, and using examples with realistic although not real data, due to confidentiality reasons that do not allow us to disclose the original application.

We present the application \app through its GUI and the corresponding test plans as designed by the company testers according to the customers' requirements, to illustrate the requirements for the automatic test generator.

\begin{figure}[ht!]
\center
\includegraphics[scale=.29]{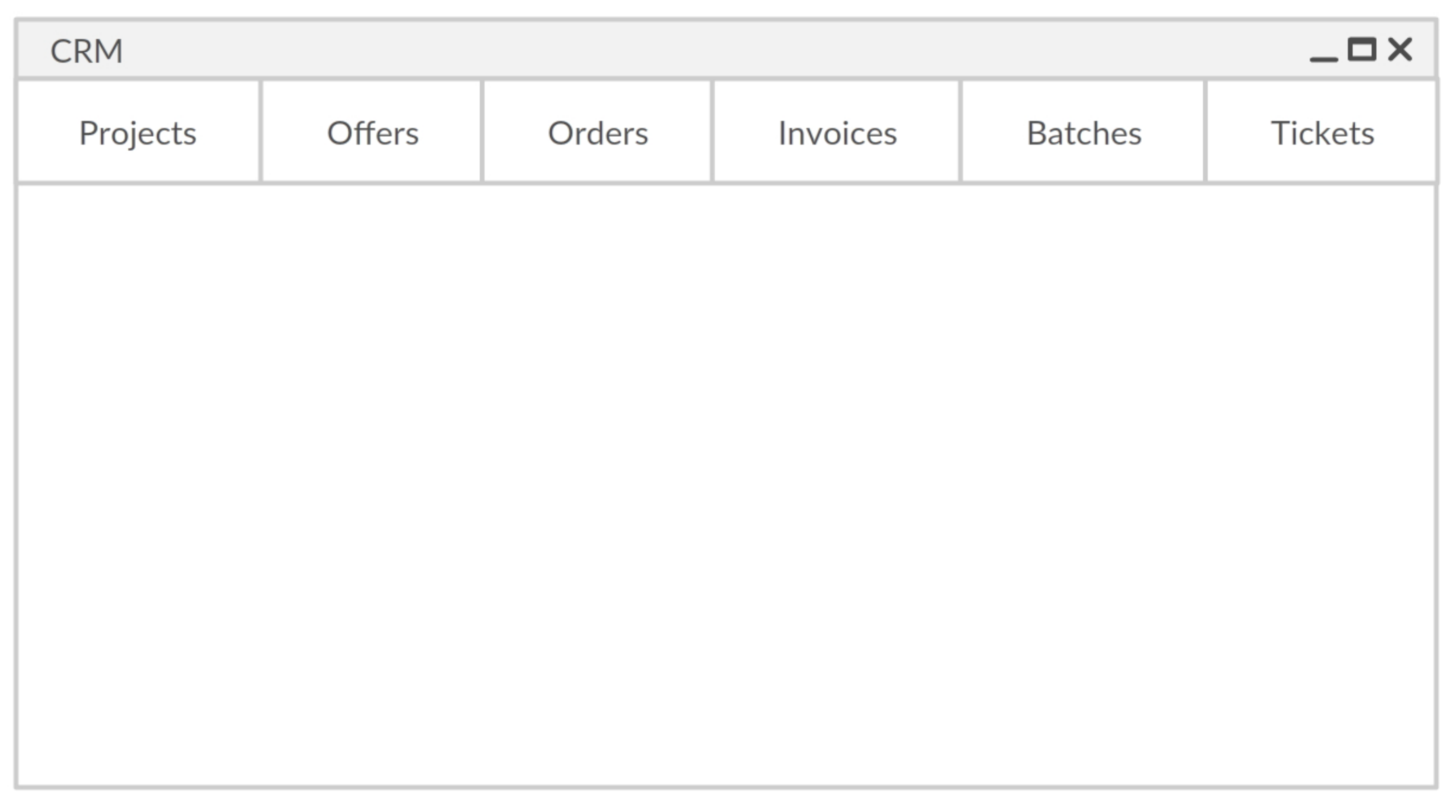}\\
(a) Home page\\~\\
\includegraphics[scale=.29]{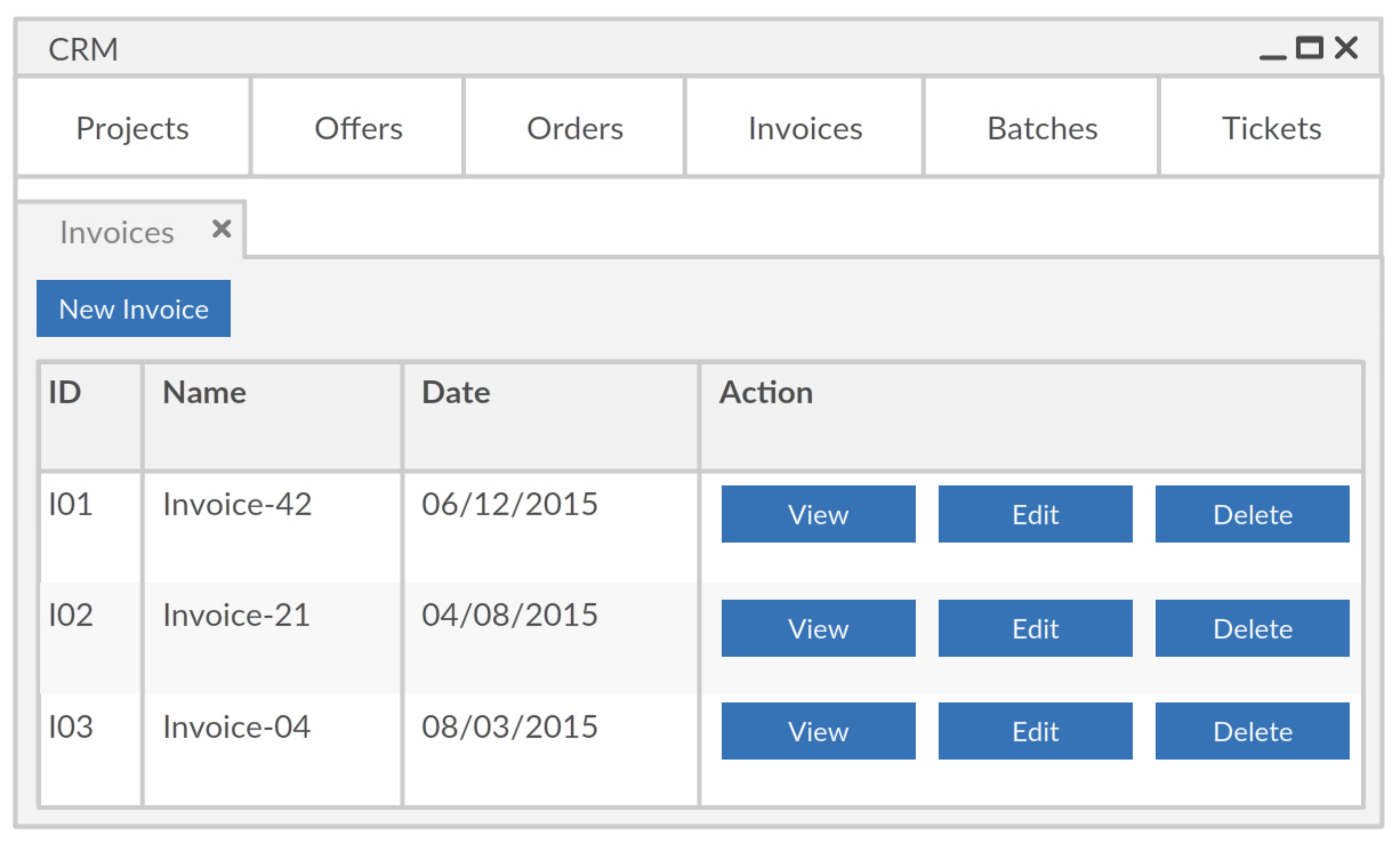}\\
(b) Invoice page\\~\\
\includegraphics[scale=.29]{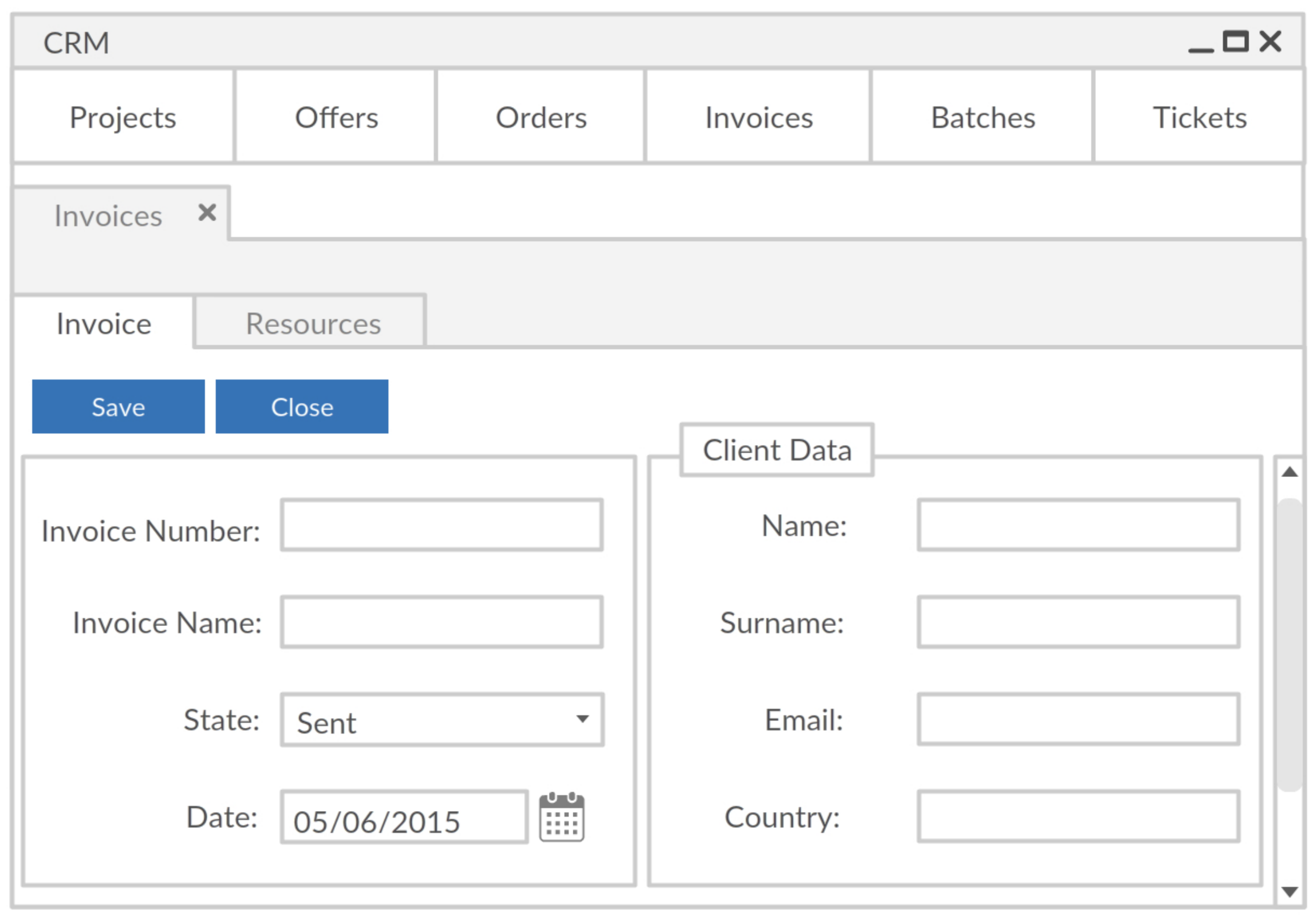}\\
(c) Edit Invoice page
\caption{GUI screens of the sample application (adapted and simplified from the original user interface)}
\label{fig:mockup}
\end{figure}

The GUI of \app is organized by data entities, following a typical interaction with the GUI, where users first select the type of entity, such as orders or invoices, and then manipulate it.
Figure~\ref{fig:mockup}(a) shows the set of six main data entities through the (simplified and anonymised) \app home page.
Buttons in the top of the home page lead to entity specific pages that offer the first tier action sets for the entity types.  
Figure~\ref{fig:mockup}(b) shows the \emph{Invoices} page with the top tier actions for entity Invoices, actions that lead to other pages when selected, for instance the \emph{Edit Invoice} page illustrated in Figure~\ref{fig:mockup}(b).
\app manages entities that include several data fields editable across multiple tabs, and editing an \app entity may require long GUI interaction sequences that can terminate successfully (\emph{Save} button) or can abort (\emph{Close} button).  \app manages six types of entities, with an average of 20 fields per entity. Editing an entity requires interacting with five tabs in average.
 
New \app releases are delivered with test plans that are composed of test objectives, where     
the test objectives are a set of behaviors exercised during testing, each consisting of  interactions executed against the target applications along with corresponding checks on the validity of the test outputs.  
Test plans are organized in sections, one for each tested entity type. 
A sample test objective for entity invoices is 

\begin{quotation}
''when the \textit{new invoice} button is pressed, a form with five tabs and only empty fields must be shown to the user''
\end{quotation}

Test plans are stored as spreadsheets with a sheet for each entity type. 
A test objective corresponds to a row in a sheet with three fields (columns): identifier, GUI interactions, and test checks.  The identifier uniquely identifies the test case. 
The GUI interactions are the set of interactions that characterise the test, for instance ``press the \textit{new invoice} button, ...". 
The test checks are a set of checks to determine the correctness of the results, for instance ``check that a form with five tabs and all empty fields is shown".

The \app test plans resemble the structure of many test plans used in small and medium size software companies that work with internally established though informally defined test procedures, and handle the test documentation with common back office tools, like spreadsheets.

\subsection{Challenges and Opportunities} 
\label{subsec:challenges}

While integrating the  test case generator \tool (AutoBlacktest~\cite{Mariani:Autoblacktest:ICST:2012}, the test generator that we described in Section~\ref{sec:abt}) in the process of our industrial partner,\footnote{
\begin{change}
Our target application, \app, is developed in C\# with Microsoft graphical libraries.
\tool supports different programming languages, including C\# and the Microsoft graphical libraries used in \app.\end{change}
} we faced both technical  and organisational challenges:
\begin{inparaenum}[(i)] 
\item the \emph{size} of the \app GUI opens a variety of interaction choices, much more than classic benchmarks used in 
\begin{change} our previous\end{change}
academic validation, with a strong impact on the effectiveness of \tool,
\item many failures derive from incorrect results that can be observed only with non trivial \emph{test oracles}, 
\item the commercial practice of the industrial partners requires \emph{test reports}   that consistently match the spreadsheet-style test plans maintained by the organization, which are not directly compatible with the outputs produced by \tool.
\end{inparaenum}
While size and oracles are technical challenges, generating test reports that meet the customers' standards is mainly an organisation-specific issue,
 \begin{change}
which 
nonetheless  must be taken into careful consideration to successfully transfer research to industry. 
\end{change}

\begin{change} These 
\end{change}
challenges generalize to many industrial applications beyond the application domain considered in this paper:
\begin{inparaenum}[(i)]
\item the difficulty of test generators to cope with huge, deep and non uniform program state spaces is well-known,
\item automatic test generators suffer from the test oracle problem, since they are seldom able to automatically determine the exact expected behavior for the functionalities tested by each test case,
\item automatic test generators often fail to meet the required test documentation standards.
\end{inparaenum}
\begin{change} 
This section contextualises the impact of these challenges in our project and discusses some opportunities that can be exploited to address them. In fact,\end{change}
though general solutions seldom exist, industrial business oriented applications can be addressed with domain specific solutions that derive from recurring design patterns, which relate the GUI structure to both the business logic and the test requirements of these applications.

\paragraph{Size of the Interaction Space} 
The classic structure of business oriented applications and the many choices that they offer at each interaction step produce an interaction space that is both extremely large and strictly structured. 

As an example at each interaction step, \app users can interact with~$95$ different menus and buttons available in the top menu bar, and with many widgets that become available in the many displayed windows.  Since the top menu bar is continuously available for interactions, the \emph{space of possible interactions grows exponentially} with the length of the interaction sequence, making a dense exploration of the execution space impossible. 
The~$95$ permanently enabled actions in the \app top bar produce a space of test sequences with at most five steps that contains more than 7 billion cases, and a several orders of magnitude larger space when considering also the actions that become available in the different windows after selecting the menu items.
Interactions sequences with more than five steps, which are common in \app, exponentially widen the already giant execution space.  
Effective test generation strategy must select a relevant subset of test cases.

\begin{change}
The main problem of testing GUIs is the huge spaces of interaction sequences that derive from the combination of windows and actions~\cite{banerjee:gui-testing:ist:2013}. Current approaches address this problem by relying on the availability of some GUI models to assist the identification of relevant interaction sequences. For example, White and Almezen concentrate on \emph{complete interaction sequences}, defined as sequences that correspond to activities of direct interest to the user, in turn defined as activities that produce observables effect on the surrounding environment of the GUI, such as changes in memory, changes in the behaviours of some peripheral devices, or changes in the underlying software or application software~\cite{white:responsabilities:issre:2000}. 
White and Almezen's approach assumes that testers provide a finite-state machine model that identifies the sets of complete interaction sequences on which testing shall focus separately. Similarly, Paiva et al.\ requires hierarchical state machines to steering the test selection strategy~\cite{paiva:hierarchical:asm:2005}. 
Yet, Memon et al.\ require testers to annotate GUI events with pre- and post-conditions, and  identify the initial and goal states that the test generator shall focus on~\cite{memon:hierarchical:tse:2001}. 
Finally, Saddler et a. also relies on pre-defined constraints to direct the test case generation process~\cite{Saddler:EventFlowSllider:ATEST:2016}.

In our project, we failed in proposing these solutions, due to the expertise, costs and time required to define the models that these approaches require. 
On the other hand, we successfully took advantage of the characteristics of the GUIs of the target applications, GUIs that are organised by data entities, as discussed in the previous section. 
We rely on the characteristic organisation of the target GUIs to automatically partition the interaction sequences into relevant and irrelevant sequences. 
The solutions that we discuss in this paper explicitly exploits the specific characteristics of the application domain, and as such, our work radically differs from the above-surveyed approaches, which generate test cases for general purpose productivity applications, like Microsoft Windows, the GVISUAL multimedia database (\cite{white:responsabilities:issre:2000}), Microsoft WordPad (\cite{memon:hierarchical:tse:2001}), Microsoft Notepad (\cite{paiva:hierarchical:asm:2005}).  
\end{change}

We defined an industrially-acceptable solution for the considered application domain, by relying on both the structured organization of the GUI and by carefully addressing operations that depend on long interaction sequences in forms with many input widgets.
The \emph{strictly structured organization of the GUI} drives the common interactions with the application, thus providing important guidelines to generate meaningful test sequences, if properly handled. 
For instance, \app users follow goal-oriented interaction patterns that correspond to the GUI organization. 
The top menus partition the interaction space according to the primary goal of the interactions,  for instance users add orders by navigating across windows enabled after selecting the order menu, and unlikely move to windows entered through top tier actions alternative to the order menu before completing the overall add operation.
Thus the execution space contains many interaction sequences that do not correspond to reasonable interactions, and test case generators that explore the execution space without considering this information produce enormous amount of almost useless test cases.

\newtcolorbox{opportunity}{
	colback=gray!5,
	colframe=gray!70,
	boxrule=0.4mm,
	left=1.5mm,
	right=1.5mm,
	top=1.5mm,
	bottom=1.5mm,
	fonttitle=\bfseries,
	title={Opportunity}
}

\medskip\begin{opportunity}
\paragraph{Graphical menus partition the functional logic} 
In most business oriented applications, top tier menus gather the operations that manipulate each type of data entities, and thus menus partition the functional logic of the application in areas.  
For example, the menus in Figure~\ref{fig:mockup} indicate that the functionalities to manipulate invoices are accessed  with action sequences that start with menu \emph{Invoices}.
This common type of GUI structure is designed to facilitate the users to easily navigate and operate on different data entities.
Test generators can exploit this structure to mitigate the combinatorial explosion of interaction sequences that the test generator may produce, by both focusing on interaction sequences that  start from distinct graphical menus, and ignoring the sequences that include actions that jump across different menus before completing any operation. 
This simple heuristic can largely reduce both the size and complexity of the execution spaces to be sampled.
\end{opportunity}

We frame our second observation  
by defining the concept of \emph{minimal interaction sequence} for a meaningful use of a system functionality as the sequence of GUI actions that must be necessarily executed to exercise a functionality relevant to the user. 
The goal of testing is to maximise the set user-relevant functionalities exercised when executing the test suite.
State-of-the-art randomized test generation strategies, \begin{change}which range from purely random strategies~\cite{Android:Monkey:WebSite,Bertolini:GuiTestingEvaluation:ICST:2010} to approaches interleaving random decisions with heuristic decisions~\cite{Mariani:Autoblacktest:ICST:2012},\end{change} 
are less and less effective in producing test cases that exercise all relevant functionalities when the length of the corresponding minimal interaction sequences grows. 
For example, state-of-the-art randomized test generation strategies can effectively generate test cases that exercise the functionalities  that correspond to minimal interaction sequences of two actions, for instance simple CRUD operations like deleting an invoice. They are much less effective in generating test cases that exercise functionalities that correspond to long minimal interaction sequences, for instance CRUD operations like modifying an invoice. 
For instance, the minimal interaction sequence for deleting an invoice consists of  selecting the menu \emph{Invoices} in the screen in Figure~\ref{fig:mockup}(a), and then clicking \emph{Delete} in the the screen in Figure~\ref{fig:mockup}(b). 
In contrast, the minimal interaction sequence for modifying an invoice consists of selecting the menu \emph{Invoices} in the screen in Figure~\ref{fig:mockup}(a), clicking \emph{Edit} in the screen in Figure~\ref{fig:mockup}(b), executing a sufficient number of \emph{fill-in} interactions with the input fields, and then clicking \emph{Save} in the  screen in Figure~\ref{fig:mockup}(c).
In a nutshell, state-of-the-art randomized test generation strategies can effectively produce test cases for the 'delete invoice' operation, but not for the 'modify invoice' operation.

To give a concrete feeling of the impact, the length of the minimal interaction sequence of operations that modify data in \app ranges from 8 to 38 steps. 
When the probability of picking up a specific action is close to 1\%, such interaction sequences are singularities in the execution space, hard if not impossible to exhaustively cover with randomized strategies. 
For instance, a ten hour run of our test generator \tool against \app generated test cases that exercised most of the short minimal interaction sequences, but hardly executed any operation whose minimal interaction sequence involved more than three actions.
In particular, \tool easily explored different selection orders of the graphical menus, but never executed the interaction sequences to modify the data entities in \app.

\medskip\begin{opportunity}
\paragraph{Long interaction sequences derive from many input widgets} 

Operations that depend on long interaction sequences, such as operations to create and modify data entities, are very challenging for automatic test case generators. 
Long interaction sequences often derive from forms with many input widgets to be filled-in. For instance, the simple input form to create a new invoice of Figure~\ref{fig:mockup} (c) includes eight input fields, and input forms with tens of input fields are common in industrial business oriented applications. 
Handling input forms as special entities can largely improve the effectiveness of test case generators, by increasing the probability of a form to be completely filled-in before being submitted. 
\end{opportunity}

\paragraph{Test Oracles} 

Effective test cases pair test inputs with relevant test oracles that can detect misbehaviours with respect to specific users' expectations.

\begin{change}
Several state-of-the-art test case generators, such as Pex~\cite{Tillmann:pex:TAP:2008}, JBSE~\cite{Braione:enhancing:esecfse:2013}, Evosuite~\cite{Fraser:whole:2013}, BiTe~\cite{Baluda:BidirectionalSymbExe:TSE:2016}, and AFL~\cite{Gutmann:afl:login:2016}, focus on computing test suites that exercise the set of executions of the applications under test as thoroughly as possible, mostly referring to code coverage metrics as indicators of testing thoroughness. 
These approaches do not generate test oracles, thus can reveal only \emph{blatant failures}, that is, uncaught exceptions and system hangs, which represent  
\end{change}
 a small fraction of failures that software companies aim to reveal with in-house testing. \begin{change2}Some techniques generate assertions to reveal regression failures in future releases of the software. Such assertions are useful for regression testing, but not for testing new and extented functionalities.\end{change2}

For example, the large majority of the test requirements in the original \app test plan include checks of the correctness of the execution that require inspecting the GUI and the database state, and cannot be identified by simply detecting uncaught exceptions and system hangs. 
Common examples of checks included in the original \app test suite are: to make sure that graphical menus, buttons, and text fields appear with correct labels and can or cannot be modified in given screens and application states; to ensure that data change operations result in correct updates of the corresponding tables in the database; to check that the application visualizes the expected subsets of data items from the many items in the database, according to  filters set in the input forms. 
Checks like these go beyond the revealing capabilities of simple implicit oracles that  reveal only uncaught exceptions and system hangs.  

\begin{change}
The problem of designing cost-effective test oracles has been largely investigated in the last decades~\cite{Barr:Oracle:TSE:2015}. 
Oracles are often provided as assertions embedded either in the test scripts, as in Junit, or in the code   in the form of contracts~\cite{Meyer:Eiffel:jss:1988,Briand:Contracts:spe:2003}.
Many approaches generate oracles from formal or semi-formal specifications, when available~\cite{Spivey:Z:Prentice-Hall:1989,Spivey:Z:Prentice-Hall:1989}.
Yet other approaches infer test oracles from informal specification in natural language~\cite{blasi:javadoc:issta:2018,Bottger:Reconciling:2001}. 
Regression testing derives oracles by monitoring the execution of the software under test~\cite{Xie:Regression:ecoop:2006}, while dynamic analysis infers invariants~\cite{Ernst:daikon:icse:1999} or models~\cite{Lorenzoli:bct:icse:2008} that can be used as test oracles.  

Unfortunately none of these approaches applied well to our context: code was given without either embedded assertions or any formal specifications, regression oracles were not sufficient, and dynamic analysis oracles were largely insufficient to represent the properties of the test plans of our industrial partner.
Nonetheless, although we could not automate the test oracles,
we mitigated the cost of manually checking the test results while executing the test suites, by automatically producing test reports with data about the effects on both the GUI widgets and database tables. These data were relevant for the manual inspection of the software requirements, and for post mortem analysis of the failures.
\end{change}

\medskip\begin{opportunity}
\paragraph{Small sets of outputs capture most relevant  classes of non-blatant failures} 

There often exists a well identified and relatively small set of output data that capture the most relevant classes of \emph{non-blatant} failures. In fact, many test requirements are concerned with checking the correctness of either attributes of widgets that frequently appear in the GUI after some operation, for instance graphical menus, buttons, text fields  and data grids, or database changes produced as a result of an operation. 
For example, an entry of the \app test plan requires to check that selecting the menu \emph{Invoices} leads the GUI to display an editing panel \emph{Invoices} that visualizes all the buttons as shown in Figure~\ref{fig:mockup} (b), and to check that filling the input form of  Figure~\ref{fig:mockup} (b) with valid data and saving it leads to inserting a new record in the database table \emph{INVOICES},  with data consistent with the data in the form. 
Thus, augmenting the output of the test generator with custom test reports that include the relevant data from both affected GUI widgets and affected database tables
could effectively assist testers in the identification of the relevant failures.
\end{opportunity}

\paragraph{Test Documentation}

In the industrial context of customised software developed on demand, such as the \app case, a fundamental requirement for the test reports is to map the test results to the test objectives in the test plans (which in turn map to the requirements). 
In fact, in many industrial sectors, software companies ask for test reports that help testers easily identify both tested and not-yet tested functionalities, to efficiently plan the test campaign and deployment plans:
Reports that include \emph{detailed information about the tested functionalities} are necessary to integrate a test generator in an industrial testing process.  

\begin{change}
The traceability between requirements and test cases is well supported when test cases are derived from the requirements~\cite{Ramesh:traceability:re:1993,Lago:Traceability:jss:2009,Winkler:traceability:sosym:2010}, but becomes a hard problem when test cases are automatically generated by sampling the input space on the implementation. 
\end{change}
State-of-the-art test case generators document the generated test suites as 
\begin{inparaenum}[(i)] 
\item executable test scripts that instantiate the test inputs, 
\item uncaught exceptions and system hangs experienced while executing the test scripts, and
 \item statement coverage. 
\end{inparaenum} 
Unfortunately, mapping the generated test cases to the system functionality by manually replaying each test script almost nullifies the benefits of using a test generator, and deducing the missed functionality by identifying inputs that lead to uncovered code can be extremely hard. 

\medskip\begin{opportunity}
\paragraph{Mapping between actions on widgets and test requirements suggest mapping from test results to system functionalities} 

The neat mapping between actions executed on the widgets in the GUI, such as clicks on graphical menus and buttons, and relevant classes of test requirements, such as validating GUI screens and database changes, provides useful information for test reports.  Pairing the actions executed on GUI widgets with relevant results on both other widgets and database tables simplifies the identification of the tested and untested test objectives, as well as the comparison between test results and test oracles.
\end{opportunity}

\bigskip 

\begin{change}
In summary,
\end{change}
our experience highlights that 
\begin{inparaenum}[(i)] 
\item graphical menus partition the functional logic of many business oriented applications,  \item long interaction sequences often derive from forms with many input widgets to be  filled-in,
\item often a small set of output data captures the most relevant classes of \emph{semantic} failures, and
\item the mapping between actions on widgets and test requirements provides useful information for mapping test results to system functionalities. 
\end{inparaenum}

 \section{GUI Exploration Strategy} 
\label{sec:strategy}

We addressed the main technical challenges of introducing automatic test case generation in industrial settings by exploiting the opportunities offered by
\begin{inparaenum}[(i)]
\item the partition of the functional logic induced by the menus defined in the GUI, 
\item the relations among action sequences induced by complex input forms, 
\item the relation between outputs and semantic faults, and 
\item the mapping between test results and system functionalities induced by the widget--test requirements mapping.
\end{inparaenum}

In this section, we present the new strategy that we defined to efficiently explore the execution space though the GUI, by exploiting the first two opportunities listed above, and discuss experimental results that confirm the effectiveness of the new strategy for the considered class of applications. Section~\ref{sec:oracles} presents the contributions related to the other opportunities.

\subsection{The SSRLS GUI Exploration Strategy}

\newtool extends \tool with a new \emph{Semi Systematic RLS (SSRLS) exploration strategy} that substitutes the \tool RLS strategy with
a systematic exploration of the GUI in two cases:

\begin{itemize}
\item \emph{Menu-driven constraints:} SSRLS constrains RLS to enforce a \emph{systematic partitioning} (\cite{Yilmaz:Combinatorial:tse:2013,Yilmaz:combinatorial:computer:2014}) of the generated test cases in groups, based on the items in the graphical menus of the GUI,  such that each group of test cases starts  accessing a graphical menu and then exercises only actions associated with the initially-accessed menu. Thus, SSRLS prevents useless and time consuming exploration of the many irrelevant interactions sequences that jump across different graphical menus.

\item \emph{Input-form constraints:} When dealing with input forms SSRLS supersedes RLS and executes complex actions that properly fill out and submit the input forms, thus increasing the coverage of the functional logic of the application under test, and avoiding the useless and time consuming exploration of too many incorrect completion of forms before correct submissions.
 \end{itemize}

In a nutshell, SSRLS combines the ability of reinforcement learning to quickly reach many different GUI states, with both the ability that systematic testing offers to optimally partition the execution space and the capability that complex actions offer to handle the implicit dependencies present in a form-based window.

Algorithm~\ref{algo:ssrls} presents the SSRLS exploration strategy by highlighting the steps added to the \tool RLS strategy:
Light grey  (steps~\ref{algo:ssrls:ifMenuAction},~\ref{algo:ssrls:firstAction},~\ref{algo:ssrls:sys1:disable} and~\ref{algo:ssrls:sys1:enable}) and dark grey background (steps~\ref{algo:ssrls:sys2:check} and~\ref{algo:ssrls:sys2:action}) highlight the statements that implement the partitioning of the execution space and the ad-hoc handling of input forms, respectively. 

The main loop (steps \ref{algo:ssrls:mainLoop:begin}--\ref{algo:ssrls:mainLoop:end}) generates an episode (a test case) per iteration until the testing budget is over. 
The algorithm initializes episodes with a sequence of actions that reach a randomly selected state of the model learned so far, that is a state observed while executing a previous episode (step~\ref{algo:ssrls:initState}). 
In particular, at the first iteration of the loop the model is empty, and the algorithm  generates an episode from the main page of the application.
In this initial state, no menu action has been executed yet, making the condition of step~\ref{algo:ssrls:ifMenuAction} evaluate to true, and the algorithm augments the episode with a menu action (step~\ref{algo:ssrls:firstAction}) followed by a sequence of GUI actions selected among the ones that become dynamically available on the GUI (loop at steps \ref{algo:ssrls:episodeLoop:begin}--\ref{algo:ssrls:episodeLoop:end}).

\begin{algorithm}[!th]
\scriptsize
\caption{\small SSRLS: the Semi Systematic RLS}
\label{algo:ssrls}
\begin{algorithmic}[1]
\Require RLS: the Reinforcement Learning based Strategy

\While{the testing budget is not over}\label{algo:ssrls:mainLoop:begin}

\State $current\_state = $ \Call{RLS}{goToARandomState}\Comment{reach a random state of the Q-learning model}\label{algo:ssrls:initState}

\If{\colorbox{light-gray}{MenuAction not done yet}} \label{algo:ssrls:ifMenuAction}
\State \colorbox{light-gray}{$current\_state = $ \Call{RLS}{doMenuAction}}\Comment{select a menu}\label{algo:ssrls:firstAction}
\EndIf

\State \colorbox{light-gray}{\Call{RLS}{disableMenuActions}}\Comment{actions on graphical menus cannot be anymore selected by RLS}\label{algo:ssrls:sys1:disable}

\For{1 $\leadsto$ number of actions per episode}\label{algo:ssrls:episodeLoop:begin}
	
	\If{\colorbox{dark-gray}{$current\_state$ is InputForm AND with some probability $\pi$}}\label{algo:ssrls:sys2:check}
	
		\State  \colorbox{dark-gray}{$current\_state =$ do\emph{FillAndSubmit}}\Comment{Fill and submit the input form}\label{algo:ssrls:sys2:action}
	
	\Else	 \State $current\_state =$ \Call{RLS}{doAction}\Comment{Execute next action}\label{algo:ssrls:rlsAction}
	\EndIf
\EndFor\label{algo:ssrls:episodeLoop:end}
\State  \colorbox{light-gray}{\Call{RLS}{enableMenuActions}}\Comment{actions on graphical menus can again be selected by RLS}\label{algo:ssrls:sys1:enable}
\EndWhile\label{algo:ssrls:mainLoop:end}
	
\end{algorithmic}

\vspace{10pt}
Support to the partitioning of the execution space (light gray background): An episode starts with a menu action and then continues only with actions  associated with the selected graphical menu.\\\\
Support to input forms (dark gray background): The  input forms are entirely filled in and submited with probability $\pi$.
\end{algorithm}

The new steps~\ref{algo:ssrls:sys1:disable} and~\ref{algo:ssrls:sys1:enable} of the algorithm disable the possibility of executing actions on the graphical menus immediately before the main \textit{for} loop, and re-enable the menu actions at the end of an episode, respectively.
These new steps guarantee that each generated test case starts with an access to a given graphical menu and exercises only actions associated with that graphical menu. If step~\ref{algo:ssrls:initState} of the algorithm executes an initial sequence of actions that already includes a menu action, step~\ref{algo:ssrls:ifMenuAction} of the algorithm prevents including additional menu actions in the current episode.
Thus, step~\ref{algo:ssrls:initState} of the algorithm can select the initial sequence of actions of an episode by navigating the current model, without any additional constraint, since the algorithm maintains the invariant that current model  contains only sequences of actions that previous episodes selected according to the SSRLS strategy.

The algorithm generates episodes of a fixed parametric length, and selects actions according to the original RSL strategy (step~\ref{algo:ssrls:rlsAction}), unless differently driven (the grey background steps that we discuss next).

Steps~\ref{algo:ssrls:sys2:check} and~\ref{algo:ssrls:sys2:action} of the algorithm control the generation of sequences of actions in the presence of input forms.
Step~\ref{algo:ssrls:sys2:check} of the algorithm checks if the current GUI state corresponds to a screen with an input form, and if so, it executes step~\ref{algo:ssrls:sys2:action} with some probability $\pi$.  Step~\ref{algo:ssrls:sys2:action} of the algorithm fills out all the input fields of the form with predefined values before clicking on the submit button.
Step~\ref{algo:ssrls:sys2:action} exploits the \tool capability to execute a set of actions. 
The probability $\pi$ is a parameter of the algorithm. 
In the experiments reported in this paper, we set $\pi =0.5$, after an empirical evaluation of different values for $\pi$.

\subsection{SSRLS effectiveness}

We extended the ABT prototype with SSRLS to validate the new approach on the industrial business oriented application made available by our project partner.  

We investigated two research questions related to the effectiveness of SSRLS:
\begin{itemize}
\item RQ1: Does SSRLS \emph{thoroughly exercise} the functionality of the business oriented application under test?

We evaluate the capability of SSRLS to generate test cases that thoroughly exercise the functionality of the application under test both in absolute terms and in comparison to RLS. We measure the effectiveness of the test cases generated with both SSRLS and RLS based on 
\begin{change}
 the frequency with which they execute different classes actions, 
with particular attention to actions that depend on long interaction sequences. 
\end{change}

\item RQ2: Does SSRLS contribute to satisfy the \emph{test objectives}?  

We evaluate the quality of the generated test cases with respect to the test plan, both in absolute terms and in comparison to RLS. We measure the significance of the generated test cases as the percentage of test objectives 
\begin{change}
(and thus corresponding requirements) 
\end{change}
that the automatically generated test cases exercise. 
\end{itemize}

We mitigated the impact of randomness by repeating each experiment five times, and reporting \begin{change2}mean values\end{change2}\begin{change}. 
We could not negotiate a higher number of repetitions with our industrial partner, due to the cost of the experiments and the limited access to the ERP application. 
The stability of results across the repeated experiments convinced us about the significance of the results in the scope of our exploratory study, but the limited amount of repetitions does not allow strong claims that our results generalise.
\end{change}

\subsubsection*{RQ1: Does SSRLS \emph{thoroughly exercise} the functionality of the business oriented application under test?}

We measure the effectiveness of SSRLS test suites  by comparing the amount of action types that the test cases generated with SSRLS and RLS exercise, respectively. We identify five main \begin{change}classes of\end{change} actions that reflect the nature of GUI interaction:

\begin{itemize}
\item \emph{Menu} actions that interact with menus; 
\item \emph{CRUD} actions that initiate a CRUD (Create, Read, Update, Delete) operation; 
\item \emph{Input} actions that enter data in input fields; 
\item \emph{SaveKO} actions that cancel submissions or submit invalid forms; 
\item \emph{SaveOK} actions that submit valid forms.
\end{itemize}  

To thoroughly exercise the functionalities of the application under test, it is important to execute both \emph{SaveKO} and \emph{SaveOK} actions, many of which are executed only with non trivial combinations of \emph{Menu}, \emph{CRUD} and \emph{Input} actions that reach specific windows and states.
We measure the effectiveness of RLS and SSRLS test cases as the amount of executed actions of different types, with specific attention to \emph{SaveKO} and \emph{SaveOK} actions. 

We study the impact of the individual optimizations introduced in this paper, by considering three configurations for SSRLS:
\begin{itemize}
\item \emph{SSRLS-partitioning} that uses only the strategy that restricts the access to the menus; it corresponds to Algorithm~\ref{algo:ssrls}, but without the statements with a dark grey background;
\item \emph{SSRLS-fillForms} that uses the complex action specifically designed to deal with forms; it corresponds to Algorithm~\ref{algo:ssrls}, but without the statements with a light grey background; 
\item \emph{SSRLS} that exploits both optimizations; it corresponds to Algorithm~\ref{algo:ssrls} with both the statements with light and dark grey background. 
\end{itemize}

\begin{figure}[ht!]
\center
\begin{scriptsize}
\includegraphics[scale=.385]{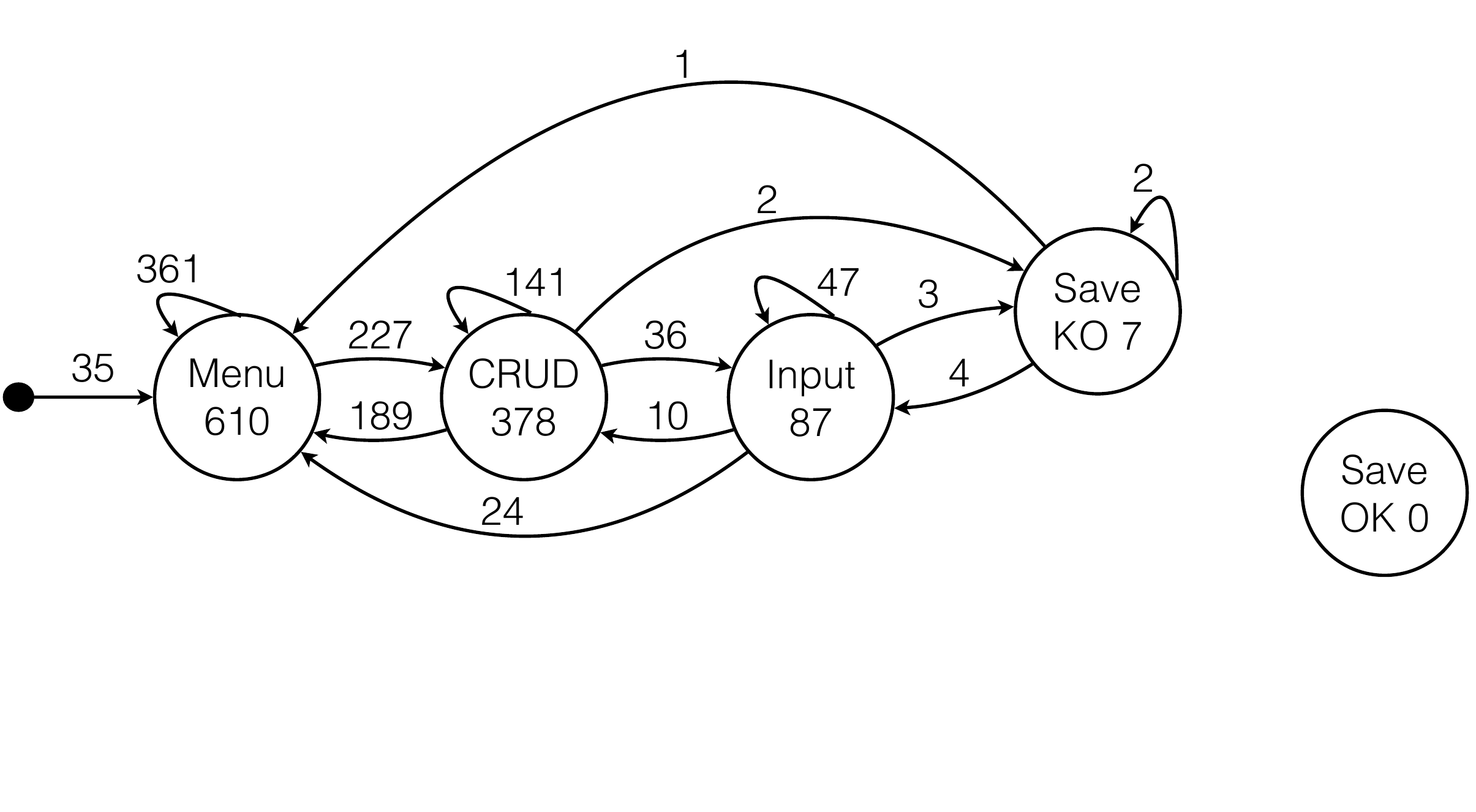}\\ 
\vspace{-50pt}
(a) RLS\\
\includegraphics[scale=.51]{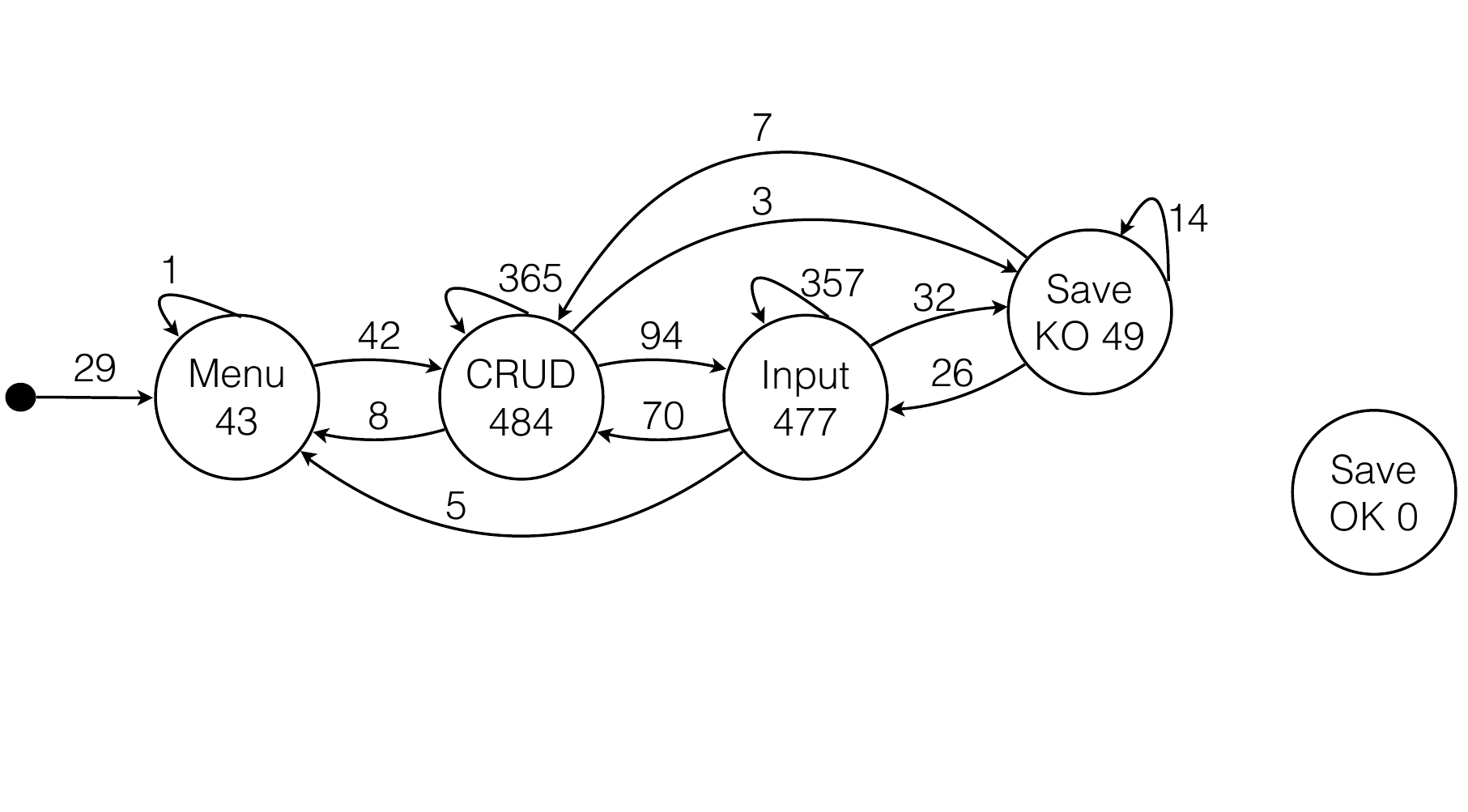}\\ 
\vspace{-50pt}
(b) SSRLS-partitioning\\
\includegraphics[scale=.385]{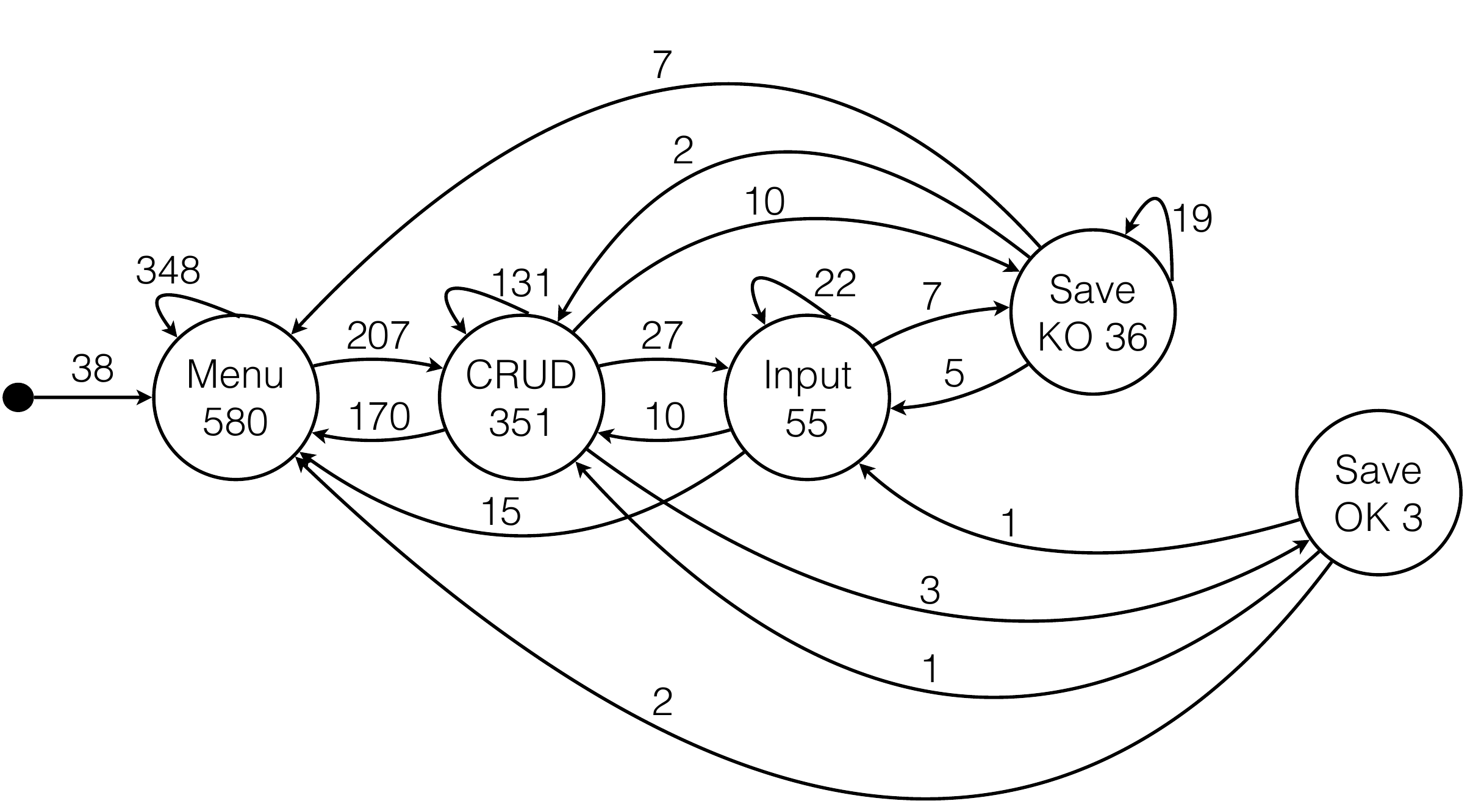}\\ 
\vspace{-5pt}
(c) SSRLS-fillForms\\
\includegraphics[scale=.51]{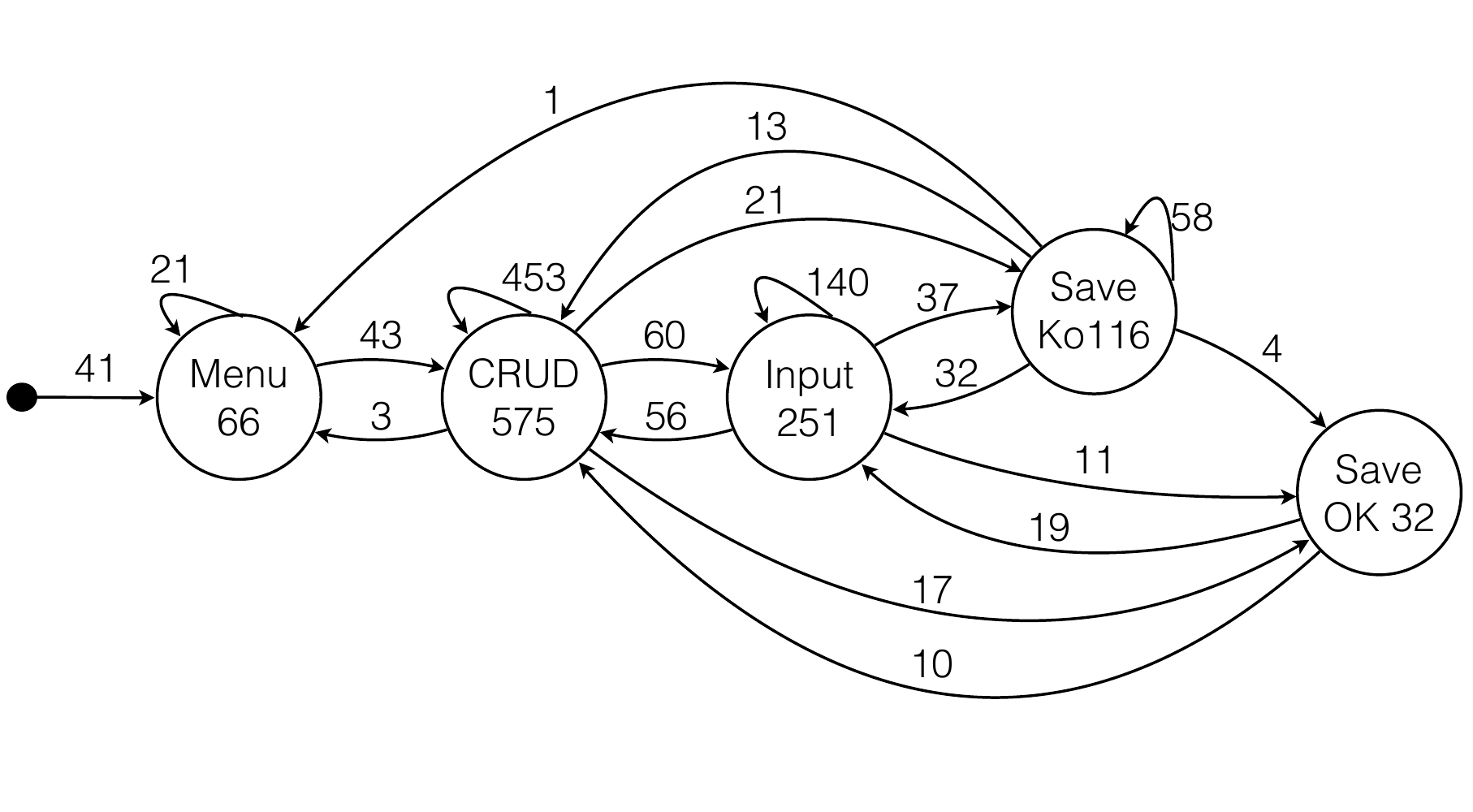}\\
\vspace{-25pt}
(d) SSRLS\\
\end{scriptsize}
\caption{Execution frequency of interaction sequences of different length}
\label{fig:interactions}
\end{figure}

We generated test cases in overnight sessions \begin{change}of six hours each, which is a practical timeframe for companies that use continuous testing solutions\end{change}, using each of the four strategies: ABT with RLS, and ABT with the three variants of the SSRLS strategy. Figure~\ref{fig:interactions} illustrates the actions executed with the different test suite as finite state models that show the sequences of GUI actions that the different test suites execute.  
The states identify the type of executed action, and the transitions represent consecutively executed action types. The weights of the states indicate how often the test cases executed that action type, and the weights of the transitions indicate how often the test suite executes consecutive action pairs.
For instance, the weight $610$ of state Menu in Figure~\ref{fig:interactions}~(a) indicates that the RSL test suite executed Menu actions $610$ times, and the weight of the transition from state Menu to state CRUD indicates that the test suite executes CRUD immediately after Menu actions $227$ times. 
\begin{change2}
Table~\ref{tab:interactions:stddev} completes the data in Figure~\ref{fig:interactions} by reporting  the mean (columns \emph{m} -- values reported in Figure~\ref{fig:interactions}) and the standard deviation (columns \emph{s}) of the execution frequency for the four considered strategies (rows \emph{RSL},  \emph{SSRLS-partitioning}, \emph{SSRLS-fillFroms}, \emph{SSRLS}) and five class of actions (columns \emph{Menu}, \emph{CRUD}, \emph{Input}, \emph{SaveOK}, \emph{SaveKO}).
\end{change2}

\begin{table}[h]
\begin{change2}
\center
\caption{Mean (m) and standard deviation (s) of the execution frequency for the  classes of actions}
\begin{small}
\begin{tabular}{l | rr | rr | rr | rr | rr}
 & \multicolumn{2}{c|}{\bf Menu} & \multicolumn{2}{c|}{\bf CRUD}&  \multicolumn{2}{c|}{\bf Input}& \multicolumn{2}{c|}{\bf SaveOK}& \multicolumn{2}{c}{\bf SaveKO}\\
\bf Strategy & m&s& m&s&  m&s& m&s& m&s\\\hline
 RLS 			& 610 & 53  & 378& 60  & 87& 31 &  0& 0  & 7& 3\\
 SSRLS-partitioning	& 43&  13  & 484& 210 & 477& 38 &  0& 0  & 49& 22\\
 SSRLS-fillForms 	& 580& 106 & 351& 103 & 55& 37 &  3& 4  & 36& 23\\
 SSRLS 			& 66& 19  & 575& 170 & 251& 105 & 32& 10 & 116& 40 \\
\end{tabular}
\end{small}
\label{tab:interactions:stddev}
\end{change2}
\end{table}

Figure~\ref{fig:interactions}~(a) shows the limitations of the RLS strategy, limitations that we discussed in Section~\ref{sec:challenges}: 
RSL test cases execute many irrelevant length 1 sequences of \emph{Menu} actions, and miss many relevant functional behaviors that depend on long interaction sequences. 
RLS test cases execute an average of $610 + 378 + 87 + 7=1,082$ actions, more than half of which (610) exercise only irrelevant length~1 sequences of \emph{Menu} actions.

Figure~\ref{fig:interactions}~(a)  also indicates that the longer it is the sequence of actions required to execute an action of a given type, the lesser RLS test suites execute that type of action. 
 RLS test suites executes \emph{CRUD} and \emph{Input} actions, which (in \app) depend on interaction sequences of length at-least~2, a mean of 378 and 87 times, respectively, \emph{SaveKO} actions, which depend on interaction sequences of length at-least~3, only 7 times, and never executes \emph{SaveOK} actions, which depend on sequences of length at-least~8.  
\begin{change2}
Table~\ref{tab:interactions:stddev} shows that the mutual strength of execution frequencies  remains stable across the considered classes of actions, despite the observation that the data related to each class of actions are subjected to some variance across the repetition of the RLS experiment.
\end{change2}

The poor results of the RLS test suites depend on the GUI structure of the subject application, a structure that is misleading for the RLS strategy that executes many actions that seldom produce significant computations. 
In fact, we observe that
\begin{inparaenum}[(i)]
\item RSL reinforcement learning assigns high reward values to interactions that cause a strong discontinuity in the GUI state, 
\item selecting one of the main data entries (top-level menu buttons) reaches windows that strongly differ from the currently visualized ones, 
\item the main data entries (top-level menu buttons) are always executable in the GUI of the subject application (see for example in Figure~\ref{fig:mockup})
\end{inparaenum}. 
As a consequence, reinforcement learning always favors selecting the main data entries (top-level menus buttons) over other actions, thus wasting most of the testing budget in interactions with the top-level menus buttons, and missing most of the application logic.

\smallskip

Figures~\ref{fig:interactions}~(b) and~(c) show the individual contributions of the two SSRLS systematic components.
SSRLS-partitioning largely reduces the time that RLS wastes in interacting with top-level menus (SSRLS-partitioning visits the Menu state $43$ times compared to $610$ RLS visits) in favour of a high rate of SaveKO Actions (SSRLS-partitioning visits the SaveKO state $49$ times compared to $7$ RLS visits). 
Although SSRLS-partitioning improves over RLS, SSRLS-partitioning test suites still miss many functionalities that require long interaction sequences: SSRLS-partitioning test suites execute \emph{SaveKO} actions that depend on interaction sequences up to length $3$, but not \emph{SaveOK} actions that depend on longer sequences.

SSRLS-fillForms also improves over RLS, but is still quite ineffective in executing actions that require long interaction sequences. 
In fact, SSRLS-fillForms test suites waste a lot of time interacting with menus (they visit the Menu state $580$ times) and rarely completes longer interaction sequences (they reach \emph{SaveKO} and \emph{saveOK} states $36$ and $3$ times, respectively). However, SSRLS-fillForms test suites improve over both RLS and SSRLS-partitioning with respect to executing \emph{saveOK} actions, thanks to being specifically designed to interact with forms. The analysis of the 3 cases in which  SSRLS-fillForms executed  \emph{saveOK} actions indicates that it
 can successfully execute operations that require interaction sequences up to length $5$ in \app. 

Figure~\ref{fig:interactions}~(d) shows that by combining the two  optimization strategies, the test suite executes many relevant application actions with a significant amount of valid and invalid cases: 
SSRLS test cases execute \emph{saveKO} and \emph{saveOK} states $116$ and $32$ times, respectively, thus showing a reasonable capability to complete long and useful interaction sequences. 
The two strategies are synergetic: the SSRLS-partitioning strategy reduces the time wasted in executing action sequences with only menu actions thus enabling the SSRLS-fillForms strategy explore long and relevant interaction sequences.
\begin{change2}
Table~\ref{tab:interactions:stddev} further supports that SSRLS is the only strategy that can steadily complete  interaction sequences that lead  to execute \emph{saveOK} actions. 

We used the (paired, one tail) Wilcoxon test to evaluate the statistical significance of the alternative hypothesis that any of the strategies SSRLS-partitioning,  SSRLS-fillForms and SSRLS improve the execution frequency of each action class with respect to RLS, based on the results of our experiments. Table~\ref{tab:interactions:wilcoxon} reports the $p$-values of each test, with reference to a considered action class (in the columns of the table) and strategy (in the rows). $p$-values lower than 0.05 (in bold in the table) indicate statistical significant improvements.
The table indicates that none of the strategies overcomes RLS with respect to the execution frequency of \emph{Menu} actions. This is expected and not particularly relevant, because the execution frequency of \emph{Menu} actions is reasonably high with any strategy, and interacting with menu items does not imply exercising relevant functionalities per se.
The relevant observation is the substantial improvement of the SSRLS strategy over RLS  with respect to the actions of all other classes, and for \emph{CRUD} and \emph{SaveOK} actions when compared to SSRLS-partitioning and SSRLS-fillForms.
\end{change2}

\begin{table}[h]
\begin{change2}
\center
\caption{Wilcoxon (paired, one tail) test}
\begin{small}
\begin{tabular}{l | r | r | r | r | r}
Alternative HP that RLS $<$ \dots & \bf Menu & \bf CRUD&  \bf Input& \bf SaveOK& \bf SaveKO\\\hline
\bf SSRLS-partitioning	& 0.98	& 0.16	& \bf 0.03	& 1	& \bf 0.03\\
\bf SSRLS-fillForms &0.71&	0.82&	0.98&	0.05&	0.05\\
\bf SSRLS 				& 1&	\bf 0.03&	\bf 0.03&	\bf 0.03&	\bf 0.03\\
\end{tabular}
\\
\begin{center}
\emph{The Wilcoxon (paired, one tail) test indicates the statistical significant values in support of the alternative hypothesis that the strategies SSRLS-partitioning,  SSRLS-fillForms and SSRLS improve on RLS for the classes of actions}
\end{center}
\end{small}
\label{tab:interactions:wilcoxon}
\end{change2}
\end{table}

\smallskip

The overall results show that SSRLS can execute not only operations that require short interaction sequences, such as deleting entities, but also operations that require long interaction sequences, such as filling-in complex form, thus significantly outperforming RLS.

\subsubsection*{RQ2:  Does SSRLS contribute to satisfy the \emph{test objectives}?  }
\label{sec:strategy:testplan:coverage} 

We measure the thoroughness of SSRLS test suites as the percentage of test objectives that test cases satisfy over the test objectives in the test plan that the testers of our industrial partner provided to us. 
The test objectives in the test plan represent the set of behaviors that are relevant to test according to the test engineers of the application.

Table~\ref{tab:testPlanCoverage} reports 
\begin{inparaenum}[(i)]
\item
the functional areas of the application (column \emph{functional area}), where a functional area is a collection of functionalities designed to handle a same type of data entities, 
\item the number of test objectives per functional area (column \emph{Test Objectives}), 
\item the number and percentage of test objectives that SSRLS and RLS test suites executed (columns \emph{Satisfied w/ SSRLS} and \emph{Satisfied w/ RLS}), respectively. 
\end{inparaenum}

The test engineers identified $350$  test objectives, the RLS and SSRLS test cases exercised 42\% and 72\% of the test objectives, respectively. This result is a clear indicator of the progresses of SSRLS over RLS. 
The higher effectiveness of the SSRLS strategy over the RLS strategy is confirmed in every functional area. 
In our project we observed that, while test engineers were reluctant in considering the original version of \tool based on RLS to automatically generate test suites, due to the relatively low coverage of the test plan that they considered insufficient to improve the overall costs of the testing tasks, they warmly welcomed \newtool that achieves reasonable functional adequacy figures  thanks to 
SSRLS.

Overall, the results of the case study show that SSRLS is significantly more effective than RLS: it fosters the exploration of deeper regions of the execution space, and generates test cases that cover a significant portion of the functional logic of the application under test. 

\begin{table}[h]
\center
\caption{Coverage of the test plan}
\begin{small}
\begin{tabular}{c | c | c | c |}
\bf Functional area & \bf Test objectives (\#) &  \bf Satisfied w/ SSRLS (\#) & \bf Satisfied w/  RLS (\#)\\\hline
Projects 	& 73		& 51 (70\%)	& 18 (25\%) \\
Orders 	& 119	& 82	(69\%) & 39 (33\%) \\
Invoices 	& 52		& 32	(62\%) & 29 (56\%)\\
Tikets 	& 21		& 20	(95\%) & 18 (90\%)\\
Modules 	& 10		& 9	(90\%) & 6 (67\%) \\
Offers 	& 75		& 57	(76\%) & 38 (51\%) \\\hline
Total 	& 350 	& 251 (72\%) & 148 (42\%)\\
\end{tabular}
\end{small}
\label{tab:testPlanCoverage}
\end{table}
 \section{Test Reports}
\label{sec:oracles}

In Section~\ref{sec:challenges} we discussed two important challenges related to the inspection of the output generated by the automatic tests: the need to support the semi-automatic validation of the test oracles specifically designed to detect relevant classes of failures in the target application domain, and the need to facilitate the evaluation of the test cases with respect to the test requirements of the applications under test. Solving these challenges is crucial to make \tool effective in industrial testing processes.

In this section, we present the new test reports that \newtool can produce and we discuss their ability to assist test analysts in the validation of the test outcomes. We first exemplify the structure of a sample test report, and discuss the design and the structure of the test reports in detail, then we present early empirical results on the effectiveness of the test reports.

\subsection{Structure and Content of the Test Reports}

\newtool generates the new test reports in tabular form for the sake of readability. The test reports indicate, for each generated test case, the actions executed on the GUI and the corresponding effect on both the GUI and the database. 
For example, Figure~\ref{fig:report} shows a test report that refers to a test case generated with 	\newtool for the application presented in Section~\ref{sec:challenges:subject}. Each row in the table represents an operation that has been executed with: 
a sequential number that identifies the operation in the scope of the test case (column \emph{ID}); 
the actions executed on the GUI (column \emph{Actions});
the effect (i.e., the changes) produced by the operation on the GUI and the database (column \emph{Outputs}).
Each individual change reported in the output starts with a tag that indicates whether the item that has been changed is an item in the GUI (tag \emph{GUI}) or in the database (tag \emph{DB}). 

In detail, the test report in  Figure~\ref{fig:report} specifies the flow of test case \emph{T3}, which starts by selecting the graphical menu \emph{Invoices} (\emph{T3.1--Actions}),  continues by clicking on the button \emph{New invoice} (\emph{T3.2--Actions}) that has appeared in the GUI as a result of the previous operation (\emph{T3.1--Outputs}), and then concludes by filling out and saving (\emph{T3.3--Actions}) the  input form visualized at the previous step (\emph{T3.2--Outputs}), which in the end causes the insertion of a new record in the database table \emph{INVOICES} (\emph{T3.3--Outputs}).

\begin{figure}[h]
\center
\begin{scriptsize}
\newcommand{\listActions}[1]{\pbox{5.5cm}{\vspace{3pt} #1 \vspace{4pt}} }
\newcommand{\listOutputs}[1]{\pbox{7cm}{\vspace{3pt} #1 \vspace{4pt}} }
\begin{tabular}{|p{.6cm} | p{5.5cm} | p{7cm} |}\hline
\multicolumn{3}{|c|}{\bf Test report of test case T3}\\\hline
\bf ID & \bf Actions  & \bf Outputs\\\hline

T3.1 & Select menu ''Invoices''  & \listOutputs{GUI: Window "Invoices" in foreground \\
								GUI: Button "New Invoices" enabled\\ 
								GUI: Grid with columns "ID", "Name", "Data", "Action" as 3 items\\
								GUI: Button "View" enabled\\
								GUI: Button "Edit" enabled\\
								GUI: Button "Delete" enabled}\\\hline

T3.2 & Click button ''New Invoice''  & 
	\listOutputs{GUI: Window "Invoice" in foreground \\
					GUI: Window "Resources" in background \\
				     	GUI: Button "Save" enabled\\ 
				     	GUI: Button "Close" enabled\\ 
					GUI: Text field ''Invoice Number'' as $\langle$empty$\rangle$ enabled\\
					GUI: Text field ''Invoice Name'' as $\langle$empty$\rangle$ enabled\\
					GUI: List field "State" ("not Sent", "Sent", "Replied")\\
					$~~~~$ as "Sent" enabled\\
					GUI: Calendar ''Date" as 05/06/2015 enabled\\
					GUI: Text field ''Client Data - Name" as $\langle$empty$\rangle$ enabled\\
					GUI: Text field ''Client Data - Surname" as $\langle$empty$\rangle$ enabled\\
					GUI: Text field ''Client Data - Email" as $\langle$empty$\rangle$\\
					$~~~~$  enabled\\
					GUI: Text field ''Client Data - Country" as $\langle$empty$\rangle$ enabled}\\\hline

T3.3 & \listActions{Fill field ''Invoice Number'' as "2015.2"\\ 
	Fill field ''Invoice Name'' as  "Payment"\\
	Fill field ''Client Data - Name'' as "Paul"\\
	Fill field ''Client Data - Surname'' as "Red"\\
	Fill field ''Client Data - Email' as\\
	$~~~~$  "paul@red.it"\\
	Fill field ''Client Data - Country'' as "Italy"\\
	Click button ''Save'' } & DB: new record in Table INVOICES $\langle$NUMBER=2015.2, LABEL=Payment, NAME=Paul, SURNAME=Red, EMAIL=paul@red.it, COUNTRY=Italy$\rangle$  \\\hline
\end{tabular}
\end{scriptsize}
\caption{A test report generated by \newtool for the case study application introduced in Section~\ref{sec:challenges}}
\label{fig:report}
\end{figure}

A test report like this one has four main qualities: 

\begin{description}
\item[Intelligible:] The descriptions reported under the column \emph{Actions} consist of labels that represent high-level domain operations. In some cases, multiple GUI actions can be grouped into a same logical Action to favour conciseness and abstraction, and produce reports that can be more easily interpreted by testers. 

\item[Easy to Validate:] The descriptions reported under the column \emph{Output} are designed to include the items that usually testers want to check to detect failures, while balancing completeness and conciseness.

\item[Use of Domain Terminology:] Test reports favour the use of domain terminology to low level technical terminology to simplify the  comprehension of the reports and the comparison of the test results with the test oracles.

\item[Efficient to Inspect:] The structure of the test report facilitates the matching between the test results and the corresponding test requirements.

\end{description}

To achieve these goals, \newtool exploits domain specific information in several ways. To generate test reports that are concise and \emph{intelligible},
\newtool groups consecutive GUI actions executed on input fields as part of a same logical operation, thus acknowledging the common characteristic of business oriented applications that rely on input forms to gather the inputs required for their operations. 
For example, with reference to the test report in Figure~\ref{fig:report}, executing the operation \emph{T3.3}, which adds a new Invoice, requires first filling in six input fields and then clicking on the \emph{Save} button. The report collapsed all these fine-grained and correlated actions into a single action of the test report.

To generate test reports that are \emph{easy to validate}, \newtool documents the changes that occur in the GUI and in the database during the execution of the test cases. These events are currently detected and reported as specified in the column \emph{Outputs} of Table~\ref{tab:outputs}. Column \emph{Source} indicates the GUI widget or the database event that may originate an entry in the test report. Column \emph{Outputs} lists the information that is extracted from the element indicated in column \emph{Source} (each variable is assigned with the value specified at the right of the $\leftarrow$ symbol). Column \emph{Format} specifies the entry that is added to the test report. Note that the entry is parametric and some values are filled in with the information extracted at runtime during test case execution.

The extracted data generally consist of the values stored on the titles and the body of the most common types of widgets, as well as the changes produced by database operations. Our experience with industrial business oriented applications suggests that the above output data is in most of the cases sufficient to validate business oriented applications against several classes of failures. The collected output data can be clearly adapted to the specific needs of an application.  

We remark that in general, as the data in the test reports are intentionally incomplete, validating some test oracles may require manual inspection of the results while replaying the generated test cases. For instance, to avoid the excessive bloating of the reports in the presence of execution states that may contain massive amounts of data, we decided to omit the content of data grids from the report and only include the titles of their columns. Oracles that check the content of the grid may require replaying a generated test case to manually inspect the grid.

To generate reports that \emph{use the domain terminology}, \newtool describes the observed outputs with domain specific terms that facilitate the comprehension of the report, as shown by the formats reported in column \emph{Format} of Table~\ref{tab:outputs}.

\begin{table}[h]
\center
\caption{Outputs in the test reports}
\begin{scriptsize}
\newcommand{\listOutputs}[1]{\pbox{5.5cm}{\vspace{3pt} #1 \vspace{4pt}} }
\begin{tabular}{l | l | p{5cm} |}

\bf Source & \bf Outputs & \bf  Format \\\hline
Graphical menu & \listOutputs{$\langle M\rangle \leftarrow$ the menu title label\\
						$\langle S\rangle \leftarrow$ the menu state \emph{enabled}/\emph{disabled}} 						& GUI: Menu $\langle M\rangle$ in state $\langle S\rangle$ \\\hline
Button &  \listOutputs{$\langle B\rangle \leftarrow$ the button title label\\
						$\langle S\rangle \leftarrow$ the button state \emph{enabled}/\emph{disabled}}   						& GUI: Button $\langle B\rangle$ in state $\langle S\rangle$\\\hline
Text field & \listOutputs{$\langle T\rangle \leftarrow$ the field title  label\\
				      $\langle I\rangle \leftarrow$ the initial value, if any, or  $\langle empty\rangle$\\
				      $\langle S\rangle \leftarrow$ the field state \emph{editable}/\emph{blocked}}   
				      & GUI: Text field $\langle T\rangle$ as $\langle I\rangle$ in state $\langle S\rangle$\\\hline
List field & \listOutputs{$\langle L\rangle \leftarrow$ the field title label\\
				      $\{\langle V\rangle\} \leftarrow$ the  possible values\\
				      $\langle I\rangle \leftarrow$ the initial value\\
				      $\langle S\rangle \leftarrow$ the field state \emph{selectable}/\emph{blocked}}   
				      & GUI: List field $\langle L\rangle$ with values $\{\langle V\rangle\}$ as $\langle I\rangle$ in state $\langle S\rangle$\\\hline
Combo-box field & \listOutputs{$\langle C\rangle \leftarrow$ the field title label\\
				      $\{\langle V\rangle\} \leftarrow$ the markable values\\
				      $\{\langle I\rangle\} \leftarrow$ the initial values, if any, or  $\langle empty\rangle$\\
				      $\langle S\rangle \leftarrow$ the field state \emph{selectable}/\emph{blocked}}   
				      & GUI: Combo-box field $\langle C\rangle$ with values $\{\langle V\rangle\}$ marked at $\{\langle I\rangle\}$ in state $\langle S\rangle$\\\hline
Data grid & \listOutputs{ $\{\langle C\rangle\} \leftarrow$ the column title labels}
				      & GUI: Grid with columns  $\{\langle C\rangle\}$\\\hline
				      
Window & \listOutputs{ $\langle W\rangle \leftarrow$ the window title label\\
				    $\langle S\rangle \leftarrow$  window in \emph{foreground}/\emph{background}}   
				    & GUI: Window $\langle W\rangle$ in $\langle S\rangle$\\\hline

Database insert & \listOutputs{ $\langle T\rangle \leftarrow$ the table name\\
				    $\langle R\rangle \leftarrow$  the inserted record}   
				    & DB: new record in table $\langle T\rangle$ as $\langle R\rangle$\\\hline
Database delete & \listOutputs{ $\langle T\rangle \leftarrow$ the table name\\
				    $\langle R\rangle \leftarrow$  the deleted record}   
				    & DB: deleted record in table $\langle T\rangle$ was $\langle R\rangle$\\\hline
Database update & \listOutputs{ $\langle T\rangle \leftarrow$ the table name\\
				    $\langle R'\rangle \leftarrow$  the record before the update event\\   
				    $\langle R\rangle \leftarrow$  the updated record}   
				    & DB: update in table $\langle T\rangle$ as $\langle R'\rangle \rightarrow \langle R\rangle$\\\hline

\end{tabular}
\end{scriptsize}
\label{tab:outputs}
\end{table}

\begin{figure}[h]
\center
\begin{scriptsize}
\includegraphics[scale=.30]{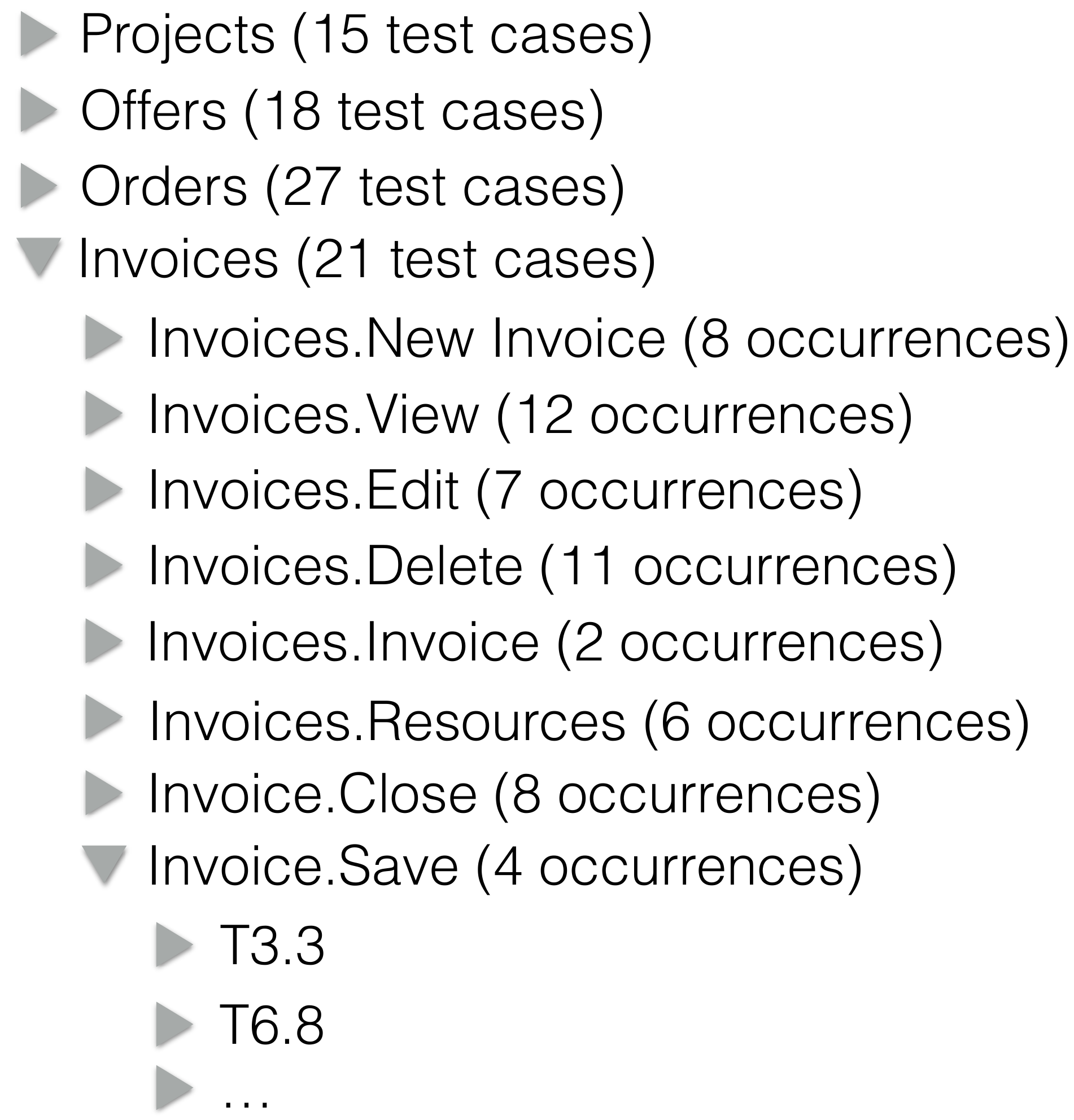}
\end{scriptsize}
\caption{The unique operation list of the test reports of \newtool: excerpt out of a test report generated for the sample application of our case study}
\label{fig:report:browser}
\end{figure}

Finally, to generate reports that are \emph{efficient to inspect}, \newtool produces browsable test reports. The browsing capability consists of the generation of a list with all the unique operations executed by the test cases included in a given test report, organized according to the graphical menus of the application. Figure~\ref{fig:report:browser} shows the unique operation list produced by \newtool for a test report generated for the sample application considered in this paper. It shows the number of test cases executed for each graphical menu, shows the number of occurrences of each operation in the test cases, and lists the individual tests that executed each operation. 

The unique operation list facilitates the tester in the inspection of the testing activity that has been automatically performed on each area of the application. 
The unique operation list in Figure~\ref{fig:report:browser}  shows that $15$, $18$, $27$ and $21$ test cases have been executed on the \textit{Projects}, \textit{Offers}, \textit{Orders}, and \textit{Invoices} menus, respectively. The operations in a same menu have been covered with different frequencies. For instance,  the operation \textit{Invoices.View} has been execute $12$ times, while the operation \textit{Invoices.Invoice} only twice. Finally, the unique operation list also indicates which operation of the test reports has executed a specific menu aciton. In the example, \textit{Invoices.Save} has been executed $4$ times in the $21$ test cases that covered the \textit{Invoices} menu. The \textit{Invoices.Save} operation occurred as the third operation of test case \emph{T3} (shown in Figure~\ref{fig:report}), as the eighth operation of the test case \emph{T6} and so forth. When clicking on the identifiers, testers are redirected to the corresponding item of the test report.

Since a test generator may generate many more test cases than the ones that could be manually inspected, the test reports tend to grow quickly. The unique operation list produced by \newtool facilitates testers in the inspection of the result produced by relevant operations, for instance by selectively inspecting the outputs produced by the execution of a same operation in many different test cases. The unique operation list also provides a concise reference to quickly identify the functional logic missed by the test generator and optimize the allocation of additional testing effort.

The implementation of the test report functionality required three main extensions to ABT: the ability to track GUI output data, the ability to track database changes, and the ability to generate browsable reports. We implemented the ability to track output data by simply adapting the GUI exploration component of \newtool. We implemented the ability to track database changes by adding a monitoring layer that relies on the standard change tracking functionality available in database servers\footnote{The interested reader may refer to the change tracking functionality of Microsoft SQL Server documented at https://technet.microsoft.com/en-us/library/cc280462(v=sql.105).aspx [Last read February 2016].}. When change tracking is active, the database server records change data in additional tables of the database, and \newtool simply retrieves data from these tables. We implemented the generation of the reports as the generation of Excel workbooks that mimic the format of the test plan that the test analysts are familiar with. The test report shows the data  that refer to distinct graphical menus of the GUI on separate spreadsheets\footnote{The possibility of formatting the reports on a per graphical menu basis is enabled by the SSRLS GUI exploration strategy described in Section~\ref{sec:strategy} that fosters the explicit mapping between test cases and graphical menus.}. 
\begin{change}
We designed the features for tracking GUI output data and database changes, and producing browsable reports, by taking advantage of the characteristics of the application domain.
Although designed for the specific domain, these features exploit common characteristics, and can be easily adapted to other environments and processes. 
Independently for the generalisability of the overall approach, some components of the approach, such as grouping actions to foster intelligibility, can be easily reused in other contexts.
\end{change}

\subsection{Effectiveness of the Test Reports}

We experienced with the test reports of \newtool while working with the industrial business oriented application made available by our project partner. In this section we discuss the empirical results on the effectiveness of the test reports.

In the light of the challenges stated in Section~\ref{sec:challenges}, we evaluate the effectiveness of the test reports by investigating the following research questions:
\begin{itemize}
\item RQ3: Do test reports \emph{suffice to verify} test oracles of business oriented applications?

\item RQ4: Do test reports facilitate \emph{scheduling additional testing activities}?
\end{itemize}

To answer RQ3, we compare the test oracles manually defined by the testers in the official test plan to the information tracked in the test reports produced by \newtool, and we quantify the portion of oracles that can be verified uniquely based on the information in the reports. 

To answer RQ4, we qualitatively discuss how the test reports that \newtool generates support the definition of additional testing tasks to increase functional coverage.

\subsubsection{RQ3: Do test reports suffice to verify test oracles of business applications?}
\label{sec:reports:testplan:coverage}
We quantify the effectiveness of the test reports by measuring the number of oracles \begin{change2}that belong to the test plan and\end{change2} that can be checked using the information in the test reports. In particular, we measure the number of \emph{verifiable oracles}, that is the number of oracles defined in the test plan that can be at least potentially verified based on automatically generated test reports, and the number of \emph{verified oracles}, that is the number of oracles that can be verified based on the actual test reports of our experiments. 

\begin{change2}As we discuss in Section~\ref{sec:challenges:subject}, the test plan classifies the test objectives by entity type. Figure~\ref{fig:test-objective-with-checks} shows an excerpt of the test plan that exemplifies a test objective for the \emph{Invoice} entities. 
A test objective corresponds to a row in a spreadsheet with three fields (columns): the identifier of the test objective, the actions that the tester shall perform to satisfy the test objective, and the checks that the tester shall perform correspondingly.  The checks represent the test oracles. In out experiments, we count each individual check as an oracle. With reference to the example in Figure~\ref{fig:test-objective-with-checks}, to satisfy the test objective \emph{8.3: Correctness of the form when adding new invoices}, the tester shall \emph{click on the button "New Invoice"}, check that (i) \emph{the foreground window is "Invoice"} and (ii) \emph{The name of the fields are} as expected. Thus, we count two oracles for this test objective. In this case, these oracles are verifiable, since they can be crosschecked on the outputs of our test reports, such as, the test report in Figure~\ref{fig:report}.
 \end{change2} 

\begin{figure}[h]
\center
\begin{scriptsize}
\newcommand{\listOutputs}[1]{\pbox{6cm}{\vspace{3pt} #1 \vspace{4pt}} }
\begin{change2} 
\begin{tabular}{|p{3cm} | p{1.5cm} | p{6cm} |}\hline
\multicolumn{3}{|l|}{\bf Test plan of application \app}\\\hline
\bf Test objective identifier & \bf Actions  & \bf Checks\\\hline
\dots & \dots & \dots\\\hline
8.3: Correctness of the form when adding new invoices & Click button "New Invoice" & \listOutputs{
- The foreground tab is "Invoice"\\
- The name of the fields are as in Table 12 of the requirements}\\\hline
\dots & \dots & \dots\\

\end{tabular}
 \end{change2} 
\end{scriptsize}
\caption{Excerpt of the test plan of \app: A sample test objective and corresponding checks (oracles)}
\label{fig:test-objective-with-checks}
\end{figure}

Table~\ref{tab:testOraclesCoverage} shows the  \begin{change2}mean\end{change2} 
results obtained for the different functional areas encompassed in the test plan. The table  reports both the number and percentage of verified oracles. The data indicate that we have been able to verify the large majority of the test oracles, that is, a total of $310$ out of the $408$ (76\%) test oracles.

\begin{table}[h]
\center
\caption{Test oracles satisfaction rates}
\begin{small}
\begin{tabular}{c | c | r  r|}
\bf Functional area & \bf Verifiable oracles (\#) &  \multicolumn{2}{c|}{\bf Verified oracles (\#)} \\\hline
Projects 	& 81		& 65 	& (80\%)	\\
Orders 	& 132	&100	& (76\%) \\
Invoices 	& 56		& 41	& (73\%) \\
Tikets 	& 38		& 28	& (74\%) \\
Modules 	& 16		& 9	& (56\%) \\
Offers 	& 85		& 67	& (79\%)  \\\hline
Total 	& 408 	& 310 & (76\%) \\
\end{tabular}
\end{small}
\label{tab:testOraclesCoverage}
\end{table}

Investigating the missed oracles in further detail, we found that 13\% ($53$ out of $408$) of the oracles could not be verified because they map on execution data that \newtool does not currently track in the reports, that is, content of data grids ($35$ oracles), graphical attributes, such as the color of GUI widgets ($6$ oracles), and database data not involved with change events ($12$ oracles). These oracles may indicate directions to extend the scope of the tracked data, though the extensions must be carefully evaluated with respect to the balance between thoroughness and conciseness, which we have discussed as an important aspect of the reports. 
The remaining 11\% ($45$ out of $408$) of the oracles express test requirements that map on execution data that do not explicitly belong to the GUI or to the database, such as verifying that an email has been sent or that some data have been written in a file. Although the test oracles that could not be checked directly using the test reports require additional testing effort to replay the test cases generated with \newtool and manually check their results, we believe that this could be an acceptable cost if limited to a small percentage of the automatic tests, as in our case where only 24\% of the test oracles cannot be checked directly from the reports. Note that the number of test cases that must be replayed is often significantly smaller than the number of test oracles that have not been checked because a single test case may be used to check multiple test oracles.

\subsubsection{RQ4: Do test reports facilitate scheduling additional testing activities}

To answer this research question, we qualitatively evaluated the cost of pairing the results in the test reports with the corresponding test requirements in the test plan, so that additional activities could be suitably defined to cover the missed items. The procedure we experienced consists of the following steps: we scan the test plan sequentially and, for each test oracle defined in the plan, we exploit the browsing capability of the test reports to identify whether the operation referred in the oracle has been executed by some test cases generated with \newtool. For each reached oracle, we further exploit the browsing capability to explode the list of test actions that executed the operation associated with the oracle, and used the first action in the list to access the output data tracked by \newtool after the execution of the test action and manually verify the oracle.

This procedure is similar to other types of methods proposed in  the scientific literature to select representative samples out of the massive amount of test results that can be produced with a test generator, for example in contexts where testers need to inspect the generated test cases. For instance, a commonly used strategy is to retain  only the test cases that cover additional branches in the code, based on the order in which the test generator computes them~\cite{Tillmann:pex:TAP:2008,Braione:TestGenerator:SwQuality:2014}. Similarly, our experimental procedure samples the test report by pairing the test results in the reports with the operations that map to test requirements in the test plan, and selecting the results that correspond to the first occurrence of each operation. 

The results reported in the paper (see Tables~\ref{tab:testPlanCoverage} and~\ref{tab:testOraclesCoverage}) indicate that, out of the test objectives and the test oracles defined in the test plan, the test cases and the test reports that \newtool generates
successfully exercize more than 70\% of the test objectives, and allow for verifying more than 70\% of the oracles related with these test objectives, respectively. 

Inspecting the test reports we have been able to schedule the manual generation of a set of test cases necessary to cover the test objectives that have not been covered automatically, and a set of automatic test cases that need to be replayed to check the test oracles could not been verified from the data in the test report. 

Overall, the test reports enabled the scheduling of new focused test activities with
significantly reduced effort. In particular, we reduced the test effort from
the  \emph{manual design, execution and inspection} of a set of test cases to cover the $350$ test objectives in the test plan, and verify the corresponding test oracles,
to:

\begin{itemize}[noitemsep]
\item the \emph{manual design, execution and inspection} of a set of test cases to cover only $99$ test objectives, since the test cases that \newtool yields automatically already exercize $251$ of the $350$ test objectives in the test plan (Tables~\ref{tab:testPlanCoverage}), 
\item the \emph{manual verification} of $98$ additional test oracles, but by means of test cases generated with \newtool: the test cases generated with \newtool hit test objectives that relate to $408$ oracles, and $310$ of these oracles can be directly verified in the test reports (Table~\ref{tab:testOraclesCoverage}), thus only  $98$ oracles require to actually replay test cases,
\item the validation of $310$ test oracles by simply browsing the test reports that \newtool produces automatically.
\end{itemize}

Based on these data, we claim that  \newtool has been able to significantly reduce the effort necessary to the design and execute the test cases, and also simplified the checking process for a large portion of the test oracles. We interprete these results as a positive indication of the possibility to optimize testing activities using a technique like \newtool.

 \section{Lessons learned}
\label{sec:results}

In this section, we report the main insights that we gained with the project activity described in this paper. We believe these insights can be helpful to drive future research effort in system testing and its application to industrial contexts. We describe our insights as a list of lesson learned.

\begin{itemize}

\item \textbf{Lesson Learned 1 - Automation is necessary but not sufficient}. Automation is highly appreciated in industry because it is a key factor to reduce development effort and costs. However, automation alone is insufficient to address the real needs of complex projects and large organizations. In particular, generating test cases that cover many of the functionalities in an application is useful only if testers can understand and interpret the testing activity that has been performed with respect to the test objectives. This is necessary to identify the functional areas that need to be further tested and checked manually (see the experience about test reports and oracles reported in Section~\ref{sec:oracles}). The only generation of tests and discovery of failures is useful but of limited value in industrial contexts, where the adequate validation of all the functionalities of an application is the priority.

\item \textbf{Lesson Learned 2 - Domain-specific approaches might scale to industrial systems} Automatically testing an application without any information about its structure and semantics is extremely challenging, and outside the capability of current automatic system testing techniques. However, we can successfully exploit domain-specific characteristics  to dramatically increase the effectiveness of testing tools and make them effective in specific domains. \begin{change}
For instance, we can use EventFlowSlicer~\cite{Saddler:EventFlowSlicer:ASE:2017} that exploits application-specific knowledge provided by testers for generating effective test cases, or Augusto~\cite{Mariani:Augusto:ICSE:2018} that exploits abstract domain knowledge to efficiently test some features of the target application.
\end{change} 
In our experience, we found useful to exploit the organization of the GUI in functional areas to make ABT more effective, up to the level of being useful in business oriented applications (see the extension to the GUI exploration strategy proposed in Section~\ref{sec:strategy}). In the future, domain specific solutions should receive greater attention from researchers and practitioners.

\item \textbf{Lesson Learned 3 - Manually-specified oracles can dramatically increase the effectiveness of automated test cases} Automatically generated test cases use implicit oracles to detect failures, that is they can only detect crashes, uncaught exceptions and hangs\begin{change}~\cite{Barr:Oracle:TSE:2015}\end{change}. The lack of powerful oracles is a major limit to failure detection. Our experience shown that the failure detection capability of the generated tests can be dramatically improved by manually specifying a few automatic oracles that can be checked at runtime (see results for RQ3). \begin{change}Our experience confirms Esbah, Deursen and Roest's results about 
the importance of human-defined  oracles to improve the effectiveness of  automated testing~\cite{Mesbah:InvariantBased:TSE:2012}.\end{change}
Although this could sometime be perceived as an extra effort for test engineers, we noticed that once a tool is perceived as useful, for instance because it can automatically cover many test objectives that had to be covered manually in the past, industry people are willing to invest their effort in the definition of program oracles that can increase the effectiveness of the synthesized test cases. Although our experience has been positive, identifying proper classes of system-level oracles that can be exploited by automatically generated test cases still deserves further research.   
 
\item \textbf{Lesson Learned 4 - Cost effective definition of system-level automated oracles is still an open challenge} While there are many languages and approaches to define unit level oracles, such as program program assertions and invariants, there is a lack of languages suitable for system level oracles. Capture and replay tools usually support the concept of checkpoint~\cite{IBR:RFT:2015}, \begin{change}  some recent studies investigate the manually specification of automatic oracles~\cite{Mesbah:InvariantBased:TSE:2012,Memon:EventFlow:STVR:2007}, but there is no complete approach to conveniently and cost-effectively specifying system-level oracles\end{change}. In fact, specifying an oracle for a test case requires either the execution of a complex sequence of interactions with the system under test or the definition of complex expressions that refer to GUI widgets. Our experience confirmed that designing system level oracles could be cumbersome. Defining more effective specification methods for the definition of system level oracles is an open challenge for the future.

\item \textbf{Lesson Learned 5 - Inflexible outputs might be a barrier to tool adoption, regardless effectiveness} In our experience, we obtained the most from ABT once it has been integrated with the testing process of the organization (\newtool). To enable the integration, it has been of critical importance to produce browsable outputs that could be exploited to guide the definition of the test strategy and plan the test effort. Without this integration, ABT would not have been considered for adoption. To make the system testing technology successful is thus important to produce solutions that can flexibly integrate with an organization process, also in terms of their input/output behavior and reporting capabilities. 

\end{itemize}

 \section{Threats to Validity}
\label{sec:threats}

\begin{change2}
The main threats to internal validity of the results reported in this paper concern the procedures to collect and analyze data. While the tool was experimented in an industrial context by professional developers and the data are the direct consequence of such activities, the authors of the paper collected and analyzed the data. To mitigate the possibility of introducing any bias in this process, we defined the procedures and the analyses as objectively as possible, severely limiting subjective judgment. The paper describes the processes to allow third parties to replicate the process for different tools and subject applications.   

Due to constraints imposed by the industrial context, it was impossible to replicate each experiment more than 5 times per configuration. Although a higher number of repetitions may generate results with higher stability, we observed an already good level of stability of the results across executions, so this limitation is not likely to affect the main conclusions of our study.

The nature of our study introduces some straightforward threads to the external validity of the results. Our goal was to investigate the adoption of \tool in an industrial scenario, and the experience reported in the paper is specific to the considered context: the company, the test case generation tools, and the  software products. Still, our experience generated interesting insights that combined with similar studies can create a useful knowledge about the challenges that must be faced when a GUI system testing technology is used in industrial projects. 
\end{change2} \section{Related Work}
\label{sec:related}

System test case generation techniques can automatically generate test cases that stimulate an application under test using its GUI. These approaches can address a range of platforms and GUI technologies, including desktop, Web, and mobile applications.

The test case generation process can be guided by different strategies. Several techniques are model-based, that is they first generate a state-based model of the GUI of the system under test and then generate test cases that cover the model according to a criterion
. The model is usually extracted by ripping the GUI of the application~\cite{Yuan:StateFeedback:TSE:2010,Xun:GIT:IEEEtrans:2011}
, and the test cases can be generated according to different criteria, such as covering sequences of events~\cite{Memon:GUIRsmoke:IEEEtrans:2005}, sequences of interacting events~\cite{Xun:GIT:IEEEtrans:2011}, or data-flow relations among event handlers~\cite{arlt:GAZOO:ISSRE:2012}.

Other techniques use search-based solutions to generate test cases that are optimal according to a given goal, for example satisfying a code coverage criterion~\cite{Gross:Exsyst:ISSTA:2012}. Other approaches simply generate test cases randomly or by combining random choices with strategies that influence randomness~\cite{Mariani:GUI:STVR:2014,Bertolini:GuiTestingEvaluation:ICST:2010}. Our research prototype, \tool, uses Q-learning to steer the testing activity, which would be random otherwise, towards the most interesting areas of the application. In this paper, we have discussed our experience with introducing  ABT
in the software process of a medium size company that develops software for third parties. We selected \tool because we wanted to avoid model-based solutions, which can generate many infeasible test cases when the functionalities under test require long interaction sequences to be executed~\cite{Bae:Comparison:JSSS:2014}, like several functionalities in our commercial system. Moreover, among the techniques that do not generate tests by covering a model, our past results suggested that ABT was an effective solution~\cite{Mariani:GUI:STVR:2014,Mariani:Autoblacktest:ICST:2012}. 

\smallskip

So far, 
only few studies considered commercial applications and the integration of automatic GUI testing solutions with the development process of a professional organization. A study that considers commercial software is the one by Bertoli et al.~\cite{Bertolini:GuiTestingEvaluation:ICST:2010} where the fault-revealing effectiveness of several GUI testing techniques is compared using Motorola smartphones as subject applications. Want et al. also empirically investigate the effectiveness of several Android testing techniques with a number commercial apps~\cite{Wang:CommercialApp:ASE:2018}. These studies focus on the comparison among techniques and do not consider the issue of introducing these techniques into the production process of an organization. Contrarily, our experience reports challenges and insights about the industrial exploitation of a GUI testing solution. Although our observations cannot be generalized to every organization and every application, and data from many other similar experiences are necessary to better understand the difficulties of introducing automatic GUI testing in industry, the experience reported in this paper represents a first step towards understanding the relationship between industry and automatic GUI testing.

\smallskip 

In our experience we faced both challenges that are well-known to the scientific community and challenges that gained little attention so far. In particular, we faced the problem of dealing with the explosion of the execution space, which is a problem present in almost every non-trivial application, but is exacerbated by the size and structure of commercial applications. We found that exploiting explicit information about the GUI and the structure of the application under test might improve the scalability of existing approaches, as reported also by Saddler and Cohen~\cite{Saddler:EventFlowSllider:ATEST:2016}. 

We also faced the oracle problem~\cite{Barr:Oracle:TSE:2015}, which is a hot research topic. While research is mostly focusing on automatically generating program oracles~\cite{Carzaniga:CrossCheck:ICSE:2014}, in our experience we realized that manually specifying the oracles might be cost-effective, but we also realized the lack of languages and approaches to cost-effectively specify them. 

Finally, the effective integration with the development process requires tools that can both produce proper outputs and suitably document the performed activity. \begin{change2}The work about automatically documenting test cases focused on the generation of code-level documentation for unit test cases~\cite{panichella:summaries:icse:2016,li:documenting:icst:2016}. In this work, we explored \end{change2}the generation of reports that can be easily interpreted by testers in terms of functional requirements that are/are not adequately tested. \begin{change2}The solution that we defined is deemed relatively cost-effective in the specific context of our experimentation. Indeed, designing automated approaches that further improve efficiency and efficacy\end{change2} is an open challenge that deserves great attention in the future.

 \section{Conclusions}
\label{sec:conclusions}

In this paper, we describe our experience in introducing a leading-edge research technology for automatically generating system tests, in a business oriented organisation, by discussing the introduction of \tool for automatically testing a business application. As a result of our experience we identify several open challenges that must be faced to effectively address large industrial applications. The most relevant ones are 
\begin{inparaenum}[(i)]
\item scalability, that is automatic system testing techniques must scale to large applications composed of many windows and functionalities that require complex input data to be executed,  
\item reporting, that is automatic system testing techniques must generate test reports that can be easily interpreted according to the test requirements of the application, and 
\item oracles, that is automatic system testing must be able to detect wrong outputs in addition to crashes.
\end{inparaenum}

Our experience indicates that it is possible to tailor effective system testing solutions to the characteristics of the application under test. 
In our industrial context, we exploited information about the structure of both the GUI and the test plan to  extended \tool to cope with the identified challenges, leading to the development of \newtool. Our results show that \newtool can reduce the effort necessary to test the system, by overcoming the main limitations that we faced with the original version of \tool. 
\begin{change}We illustrate the elements that led us identifying and implementing the improvements needed for introducing the approach in production, and discuss the open challenges towards a routinely approach for automating the generation of system-level test suites. \end{change}

\begin{change}
An exploratory case study is a preliminary step toward general solutions, which serves to generate research hypotheses for specific and focused causal research.
Thus, we do not claim that our current results directly generalise to all contexts. 
Nonetheless, the experience that we report in this paper provides important empirical evidence about effective automation of system testing in industrial settings.
The industrial study that we discuss in the paper indicates that oracles, efficient exploration of the interaction space, and generation of useful reports are critical enabling factors for transferring leading-edge approaches for automatic test case generation into the production line. 
The solutions that we developed to address these issues pave the way towards architecting
industrial-strenght system test case generation approaches.
\end{change}

\begin{change}
Currently our industrial partner can autonomously run \newtool on the ERP application considered in the study, even though they  can hardly address new applications without our support, due to the difficulty in setting up new application-specific configurations and deploying the tool in the context of new software projects. 
Based on the results of the experience that we discuss in this paper, 
we are now working with an industrial partner to productise \newtool, which is still in a prototype stage and needs to be properly embedded in a commercially usable solution. 
\end{change}

\begin{change}
Our current research agenda aims to study new solutions towards tailorable system testing approaches that can be easily adapted to the characteristics of specific classes of applications. 
We are collaborating with new industrial partners to collect additional evidence of the effectiveness of the approach discussed in this paper with new business oriented applications.
We aim to investigate the practicality of the tool and the actionability of the generated outputs for different scenarios, aiming to assess the general validity of the results reported in this paper.
\end{change} 
We are also actively conducting research on developing effective automatic test generation approaches that address scalability, reporting and oracles. 
\section*{Acknowledgements} 
This work has been partially supported by  the H2020 ERC Proof of Concept project AST (grant agreement n. 824939) and by the SISMA national research project (MIUR, PRIN 2017, Contract 201752ENYB).

\bibliographystyle{wileyj}

\end{document}